\documentclass[apj]{emulateapj}

\usepackage{graphicx,amssymb,amsmath,amsfonts}
\usepackage{color,subfigure}
\usepackage{natbib}
\usepackage[section]{placeins}
\usepackage{cases}
\usepackage[mathscr]{euscript}
\usepackage{mathrsfs}

\usepackage{gensymb}

\newcommand{\appropto}{\mathrel{\vcenter{
  \offinterlineskip\halign{\hfil$##$\cr
    \propto\cr\noalign{\kern2pt}\sim\cr\noalign{\kern-2pt}}}}}

\def\app#1#2{%
  \mathrel{%
    \setbox0=\hbox{$#1\sim$}%
    \setbox2=\hbox{%
      \rlap{\hbox{$#1\propto$}}%
      \lower1.1\ht0\box0%
    }%
    \raise0.25\ht2\box2%
  }%
}

\renewcommand{\th}{$^{\rm th}$}

\ifdefined\kms
\renewcommand{\kms}{\,\rm km\ s^{-1}}
\else
\newcommand{\kms}{\,\rm km\ s^{-1}}
\fi

\let\AAold\AA
\renewcommand{\AA}{\text{\AAold}}
\newcommand{\mic}{\,\mbox{$\mu$m}}
\newcommand{\cm}{\,{\rm cm}}

\newcommand{\pc}{\,{\rm pc}}
\newcommand{\kpc}{\,{\rm kpc}}
\newcommand{\Mpc}{\,{\rm Mpc}}

\newcommand{\ryd}{\,{\rm Ryd}}

\newcommand{\yr}{\,{\rm yr}}
\newcommand{\Myr}{\,{\rm Myr}}

\newcommand{\g}{\,{\rm g}}
\newcommand{\K}{\,{\rm K}}

\newcommand{\msun}{\,{\rm M_{\odot}}}
\newcommand{\zsun}{\,{\rm Z_{\odot}}}

\renewcommand{\mp}{m_{\rm p}}

\newcommand{\deriv}{{\rm d}}

\newcommand{\dex}{\,{\rm dex}}

\DeclareRobustCommand{\ion}[2]{%
\relax\ifmmode
\ifx\testbx\f@series
{\mathbf{#1\,\mathsc{#2}}}\else
{\mathrm{#1\,\mathsc{#2}}}\fi
\else\textup{#1\,{\mdseries\textsc{#2}}}%
\fi}

\newcommand{\hi}{\text{H~{\sc i}}}

\newcommand{\niip}{\text{N~{\sc ii}}}
\newcommand{\nvp}{\text{N~{\sc v}}}

\newcommand{\Siiip}{\text{Si~{\sc ii}}}
\newcommand{\Siiiip}{\text{Si~{\sc iii}}}
\newcommand{\Siivp}{\text{Si~{\sc iv}}}

\newcommand{\oip}{\text{O~{\sc i}}}

\newcommand{\ovip}{\text{O~{\sc vi}}}

\newcommand{\neviiip}{\text{Ne~{\sc viii}}}

\newcommand{\mgip}{\text{Mg~{\sc i}}}
\newcommand{\mgiip}{\text{Mg~{\sc ii}}}

\newcommand{\mgxp}{\text{Mg~{\sc x}}}

\newcommand{\ciiip}{\text{C~{\sc iii}}}
\newcommand{\ciip}{\text{C~{\sc ii}}}
\newcommand{\civp}{\text{C~{\sc iv}}}

\newcommand{\Lya}{\text{Ly$\alpha$}}

\ifdefined\aap
\else
\newcommand{\aap}{A\&A}
\newcommand{\araa}{ARA\&A}
\newcommand{\apjl}{ApJ}
\newcommand{\apjs}{ApJS}
\newcommand{\apj}{ApJ}
\newcommand{\aj}{AJ}
\newcommand{\mnras}{MNRAS}

\fi
\newcommand{\rmxaa}{RevMexA\&A}

\newcommand{\cloudy}{{\sc cloudy}}

\newcommand{\aion}{\alpha_{\rm ion}}

\newcommand{\nH}{n_{\rm H}}
\newcommand{\NH}{N_{\rm H}}

\newcommand{\rhobar}{{\bar \rho_b}}
\newcommand{\Mhalo}{M_{\rm halo}}
\newcommand{\NHI}{N_{\rm HI}}

\newcommand{\fion}{F_{\rm ion}}

\newcommand{\AMD}{\deriv \NH / \deriv (\log \rho)}
\newcommand{\AMDfrac}{\frac{\deriv \NH}{\deriv \log \rho}}
\newcommand{\nhi}{N_{\rm \hi}}
\newcommand{\novi}{N_{\rm \ovip}}

\newcommand{\nsiiv}{N_{\rm \Siivp}}

\newcommand{\nnii}{N_{\rm \niip}}

\newcommand{\rvir}{R_{\rm vir}}

\newcommand{\dNz}{N_0}
\newcommand{\nz}{n_{{\rm H},0}}
\newcommand{\rz}{r_{c,0}}

\newcommand{\phii}{\phi}
\newcommand{\phiHM}{\phi_{{\rm HM12}}}

\newcommand{\Rimp}{R_{\perp}}
\newcommand{\Nexp}{N_{\rm pred}}
\newcommand{\Nobs}{N_{\rm obs}}
\newcommand{\CF}{f_{\rm C}}

\newcommand{\Mcool}{M_{{\rm cool}}}

\newcommand{\rc}{r_{\rm c}}
\newcommand{\nc}{n_{\rm c}}
\newcommand{\sigc}{\sigma_{\rm c}}
\newcommand{\rci}{r_{{\rm c},i}}

\newcommand{\fV}{f_{{\rm V}}}
\newcommand{\fVz}{f_{{\rm V},0}}

\newcommand{\nHi}{n_{{\rm H},i}}

\newcommand{\rhoi}{\rho_{i}}
\newcommand{\rhoim}{\rho_{i-1}}
\newcommand{\rhomin}{\rho_{\rm min}}

\newcommand{\rhomax}{\rho_{\rm max}}

\newcommand{\NHavRimp}{\langle\NH(\Rimp)\rangle}

\newcommand{\nHbar}{{\bar \nH}}
\newcommand{\el}{{\rm X}}
\renewcommand{\ion}{\el^{i+}}
\renewcommand{\fion}{f_{\ion}}
\newcommand{\Nion}{N_{\ion}}
\newcommand{\rhoz}{\rho_0}
\newcommand{\rhocool}{\rho_{\rm cool}}

\newcommand{\Thot}{T_{\rm hot}}

\newcommand{\fg}{f_{\rm gas}}

\begin{document}

\title{A universal density structure for circum-galactic gas}
\author{
Jonathan~Stern\altaffilmark{1}\footnotemark[*]\footnotemark[\textdagger], 
Joseph~F.~Hennawi\altaffilmark{1},
J.~Xavier~Prochaska\altaffilmark{2}, and
Jessica~K.~Werk\altaffilmark{2,3}
}
\footnotetext[*]{E-mail: stern@mpia.de}
\footnotetext[\textdagger]{Alexander von Humboldt Fellow}
\altaffiltext{1}{Max Planck Institut f\"{u}r Astronomie, K\"{o}nigstuhl 17, D-69117, Heidelberg, Germany} 
\altaffiltext{2}{UCO/Lick Observatory; University of California, Santa Cruz, CA}
\altaffiltext{3}{Astronomy Department at the University of Washington, Seattle, WA}
\begin{abstract} 
We develop a new method to constrain the physical conditions in the cool ($\sim10^4\K$) circumgalactic medium (CGM) from measurements of ionic column densities, by assuming that the cool CGM spans a large range of gas densities and that small high-density clouds are hierarchically embedded in large low-density clouds. 
The new method combines the information available from different sightlines during the photoionization modeling, thus yielding tighter constraints on CGM properties compared to traditional methods which model each sightline individually.  
Applying this new technique to the COS-Halos survey of low-redshift $\sim$\,$L^*$ galaxies, we find that we can reproduce all observed ion columns in all 44 galaxies in the sample, from the low-ions to $\ovip$, with a single universal density structure for the cool CGM. 
The gas densities span the range $50\lesssim\rho/\rhobar\lesssim5\times10^5$ ($\rhobar$ is the cosmic mean), while the physical size of individual clouds scales as $\sim\rho^{-1}$, from $\approx35\kpc$ of the low density \ovip\ clouds to $\approx6\pc$ of the highest density low-ion clouds. 
The deduced cloud sizes are too small for this density structure to be driven by self-gravity, thus its physical origin is unclear. 
The implied cool CGM mass within the virial radius is $(1.3\pm0.4)\times10^{10}\msun$ ($\sim$1\%\ of the halo mass), distributed rather uniformly over the four decades in density. The mean cool gas density profile scales as $R^{-1.0\pm0.3}$, where $R$ is the distance from the galaxy center.
We construct a 3D model of the cool CGM based on our results, which we argue provides a benchmark for the CGM structure in hydrodynamic simulations.
Our results can be tested by measuring the coherence scales of different ions.
\end{abstract} 


\section{Introduction} 	

Observations of the circumgalactic medium (CGM), 
defined loosely as gas within the halo virial radius $\rvir$ but outside the galaxy main stellar body, 
can constrain two crucial processes in the formation of galaxies -- 
inflows from the intergalactic medium (IGM) and outflows from the galaxy.
The CGM is also a potential site for some of the `missing baryons',
which are baryons expected from big bang nucleosynthesis but unaccounted for by observations (\citealt{Fukugita+98, Bell+03}). 
Therefore, estimates of the CGM mass and its physical properties provide important constraints for both theories of galaxy formation, and for the inventory of cosmic baryons.

The mass of the cool ($T\sim10^4\K$) baryons in the CGM, $\Mcool$, can be derived
from an estimate of the average photoionized hydrogen column $\langle\NH\rangle$
through the CGM, via
\begin{eqnarray}\label{eq: M vs N}
 \Mcool & \sim & \pi \rvir^2 \frac{\mp}{X} \langle\NH\rangle \nonumber \\
 & \approx & 3\times 10^{10} \frac{\langle\NH\rangle}{10^{19}\cm^{-2}} \left(\frac{\rvir}{300\kpc}\right)^2 \msun ~,
\end{eqnarray}
where $\mp/X\approx1.4\mp$ is the gas mass per hydrogen particle.
A possible approach to estimate the average $\NH$ is to compile a sample of projected galaxy-QSO pairs without any absorption pre-selection. 
Then for each galaxy, one can measure the ionic column densities along the sightlines to the background quasar, 
and apply an ionization correction in order to deduce the total column. 
This approach yields a relatively unbiased census of the gas around galaxies.
Recently, \cite{Werk+14} applied this method to the COS-Halos survey of $\sim L^*$ galaxies at redshift $z\sim 0.2$ (\citealt{Thom+12, Werk+12,Werk+13, Tumlinson+13}), and found $\Mcool \gtrsim 6.5\times 10^{10}\msun$, 
more than the typical stellar mass in the galaxies in their sample. 
Additional similar surveys have been undertaken in order to estimate the CGM properties of galaxies with different luminosities and redshifts
(\citealt{Hennawi+06a,ProchaskaHennawi09, Crighton+11, Rudie+12,Prochaska+13,Bordoloi+14,Lau+15}). 

Most studies of the CGM assume that the absorption features come from gas with some characteristic volume density $\rho$, and therefore some characteristic ionization level ($\propto \rho^{-1}$).
However, when $\rho$ is optimized to reproduce the column of low-ionization ions such as \Siiip\ and \Siiiip, the observed \ovip\ columns are underpredicted by orders of magnitude (e.g.\ \citealt{Werk+14}). 
In some cases, even the column of the lower ionization \Siivp\ is underestimated by the single-density models optimized to fit the low ions (Werk et al.\ 2016).
When higher-ionization ions such as \neviiip\ are observed, their observed columns are also severely underpredicted by the single-$\rho$ models fit to the low-ions (\citealt{Savage+05b, Narayanan+11, Meiring+13}). These discrepancies suggest that single-$\rho$ models are likely an oversimplification, and have led the authors of these studies to argue for a multi-phase CGM. 
However, once multiple phases are invoked, our ability to observationally constrain the CGM properties drops considerably, due to the extra free parameters.
For example for the high-ion phase, which is typically traced only by \ovip\ since other high-ions such as \neviiip\ are challenging to observe, both photoionization and collisional ionization have been invoked (e.g.\ \citealt{Savage+02}), though neither can be ruled out. This uncertainty in the ionization mechanism
results in huge uncertainties in the mass of this phase and hence in the total CGM mass (\citealt{Tumlinson+11, Peeples+14, Werk+14}).  
Another disadvantage of the multi-phase picture is that it does not naturally explain why the kinematics of \ovip\ and other high ions are commonly found to be aligned with the kinematics of the
low-ions (e.g.\ \citealt{Simcoe+02,Simcoe+06,Prochaska+04,Tripp+11,Fox+13,Werk+14,Crighton+15}).

In this paper we introduce a new method to model circumgalactic photoionized gas which spans a range of densities, under the assumption that the different densities are spatially associated as suggested by the line kinematics. Our new method
uses absorption line modeling to derive the
hydrogen column per decade in density $\AMD$, which is the natural extension of the total column $\NH$ to multi-density gas. 
We show below that the $\AMD$ formalism allows combining information on the density structure from different objects 
during the absorption line modeling, i.e.\ it allows one to model the `stacked' CGM of a large ensemble of observations of different galaxies. This stacking yields tight constraints on the properties of the multi-density CGM,
compared to traditional methods in which each object is modeled individually, and the aggregate CGM properties are deduced from some average over the individual absorption models.

The quantity $\AMD$, known as the Absorption Measure Distribution (AMD), is the absorption analog of the emission measure distribution (EMD) widely used in the analysis of emission-line spectra.
Its importance was recognized in the context of `warm absorbers' -- outflowing gas seen as absorption features in X-ray spectra of Active Galactic Nuclei.
Analysis of the AMD
in these systems was used both to demonstrate the existence of a thermal instability in the absorbing gas, and to constrain the physical conditions in the outflows (\citealt{Holczer+07, Blustin+07, Behar09, HolczerBehar12, Stern+14b, Adhikari+16, Goosmann+16}).
For the CGM, equation~(\ref{eq: M vs N}) suggests that an estimate of $\AMD$ yields a constraint on $\deriv \Mcool / \deriv (\log \rho)$, namely the cool gas mass distribution as a function of gas density. This mass distribution can be directly compared to the predictions of hydrodynamic simulations.

This paper is structured as follows. 
In \S\ref{sec: method} we present the formalism we use to analyze a multi-density CGM, 
and describe our method to derive $\AMD$ in a sample of galaxy-selected absorbers. In \S\ref{sec: application} we apply this method to the COS-Halos sample, 
while in \S\ref{sec: CGM characteristics} we use the determined $\AMD$
to deduce aggregate characteristics of the CGM of COS-Halos galaxies. 
We discuss the uncertainties and implications of our results in \S\ref{sec: discussion1}.
In \S\ref{sec: conclusions} we summarize our results and suggest how they can be expanded in future work.

\section{Formalism and Method}\label{sec: method}

In this section we provide analytic estimates for the absorption features expected in a multi-density cool CGM, followed by a prescription for calculating a more accurate solution using the numerical photoionization code \cloudy\ (\citealt{Ferland+13}).  
The formalism is based on the formalism in \cite{HennawiProchaska13}, who discussed a CGM filled with cool clouds which have the same density and size, and are distributed uniformly within the virial radius $\rvir$. We first generalize the \citeauthor{HennawiProchaska13} formalism to allow a cloud distribution which varies as a function of the distance from the galaxy $R$, and then further generalize to a CGM with clouds with different densities.

\subsection{Single-Density Cool Cloud Model}

A CGM filled with homogeneous spherical clouds can be 
characterized using the cloud density $\rho$, the cloud size $\rc$, and the volume filling factor $\fV$. 
We assume a spherically-symmetric CGM in which $\fV$ varies as a power law in $R$, and is zero beyond $\rvir$:
\begin{equation}\label{eq: fV def single rho}
\fV(R) = \fV(\rvir) \left(\frac{R}{\rvir}\right)^{l} ~~ \{R < \rvir\}  
\end{equation}
The above five parameters $(\rvir,\rho,\rc,\fV(\rvir),l)$ define this idealized cool CGM model, and can be used to derive aggregate CGM quantities. The cool CGM mass within $R$ 
is (for $l>-3$) 
\begin{eqnarray}\label{eq: Mcool single rho}
 \Mcool(<R) &=& \int_{0}^{R}4\pi R^2 \rho \fV(R)\deriv R \nonumber\\
            &=& \frac{4\pi R^3\rho\fV(\rvir)}{3+l}    \left(\frac{R}{\rvir}\right)^{l} ~.
\end{eqnarray}

The above five parameters can also be used to calculate quantities which are either directly observable or closely related to absorption line observations. One such quantity is the average column density $\langle\NH\rangle$, where we use the symbol $\langle \cdot \rangle$ to denote an average over an ensemble of sightlines through the CGM of a galaxy, or an ensemble of sightlines through the CGM of a homogeneous selected sample of galaxies.
The value of $\langle\NH\rangle$ as a function of impact parameter $\Rimp$ is related to $\fV$ and $\rho$ via 
\begin{equation}\label{eq: NH Rimp avg single rho}
\NHavRimp = \nH \int \fV(R)\deriv s 
\end{equation}
where $\nH = X\rho/\mp$ is the hydrogen number density, and $\deriv s$ is the line element. The integral in eqn.~(\ref{eq: NH Rimp avg single rho}) is equal to the average pathlength of the sightline through the cool clouds. This integral is further developed in 
\S\ref{sec: CGM characteristics} where we constrain $\fV$ from observational data.

A second observable is the covering factor $\CF(\Rimp)$, defined as the chance a line of sight intersects at least one cloud\footnote{This is a somewhat different definition then used by \cite{HennawiProchaska13}, who defined $\CF$ as the average number of clouds along the sightline. The two definitions are equivalent if cloud overlap along the sightline is negligible.}.
The relation between $\CF$, $\fV$ and $\rc$ is straightforward in the limit that clouds do not overlap along a single line of sight. 
In this limit, the contribution to $\CF$ per unit length is $\deriv\CF/\deriv s=\nc\sigc$, 
where $\nc = \fV/(4\pi \rc^3/3)$ is the cloud number density 
and $\sigc=\pi \rc^2$ is the cloud cross-sectional area. Hence
\begin{equation}\label{eq: CF Rimp single rho}
 \CF(\Rimp) = \int \nc(R)\sigc \deriv s = \frac{3}{4\rc}\int \fV(R)\deriv s  
\end{equation}
Using eqns.~(\ref{eq: NH Rimp avg single rho}) and (\ref{eq: CF Rimp single rho}) to solve for $\rc$ we get
\begin{equation}\label{eq: rc single rho}
 \rc = \frac{3\langle\NH(\Rimp)\rangle}{4\nH\CF(\Rimp)} ~.
\end{equation}

Therefore, in the context of this single-density cloud model one can constrain $\rho$, $\CF(\Rimp)$ and $\NHavRimp$ from the observations, and then use equations~(\ref{eq: NH Rimp avg single rho})--(\ref{eq: rc single rho}) to deduce the physical parameters $\rc$, $\fV(\rvir)$, and $l$. 
These parameters can then be used to derive the aggregate characteristics of the CGM. 

It is important to note that this formalism allows calculating only average observational quantities, because a single sightline depends on a specific realization of the CGM, and the the resulting stochasticity is not fully specified by the five parameters mentioned above.

\subsection{Multi-Density Cool Cloud Model}\label{sec: multi-rho}

We now generalize the model from the previous section to CGM clouds which span a range of gas densities. For the analytic formalism we utilize a discrete picture in which CGM clouds can have one of a set of densities $\rhoi$, in which consecutive values differ by an order of magnitude, i.e.\ $\rhoi=10\rhoim$. 
We find this discrete picture conceptually and notationally simpler than a more realistic scenario where $\rho$ varies continuously. This discrete picture is also used for visualization purposes.
However, for increased accuracy in the numerical photoionization calculation below, we use the finer sampling in $\rho$ used by \cloudy, where consecutive values of $\rho$ typically differ by $\Delta\log\rho\approx0.01$.\footnote{\cloudy\ divides the calculated slab into layers, where the depth of each layer is chosen such that the physical conditions are roughly uniform across the layer.}

We assume that all clouds with a given density have the same size 
\begin{equation}\label{eq: density structure}
 \rc(\rhoi) = \rz \left(\frac{\rhoi}{\rhoz}\right)^\alpha ~,
\end{equation}
where $\rhoz$ is the lowest density in the cool CGM and $\rz\equiv\rc(\rhoz)$. 
The filling factor of each cloud type is assumed to have the form
\begin{equation}\label{eq: fV def}
\fV(\rhoi,R) = \fVz \left(\frac{\rhoi}{\rhoz}\right)^m \left(\frac{R}{\rvir}\right)^l ~,
\end{equation}
i.e.\ the dependence of $\fV$ on $\rho$ is assumed to be separable from the dependence of $\fV$ on $R$. More complicated forms for $\fV$ are not well-constrained with the COS-Halos sample used below, and are not analyzed in this work. 

Using these seven parameters $(\rvir,$ $\rz$, $\rhoz$, $\alpha$, $\fVz$, $m$, $l)$, it is straightforward to generalize the equation for the mass within $R$ (eqn.~\ref{eq: Mcool single rho}) to a multi-density CGM: \begin{equation}
\label{eq: Mcool}
 \Mcool(<R) = \frac{4\pi R^3\rhoz\fVz}{3+l}  \left(\frac{R}{\rvir}\right)^{l} \sum_{\rhoi} \left(\frac{\rhoi} {\rhoz}\right)^{m+1} ~.
\end{equation}
Another interesting property is the distribution of cool gas mass within $\rvir$ as a function of gas density, which is equal to
\begin{equation}
\label{eq: dM/drho}
 \frac{\deriv\Mcool}{\deriv\log \rho}(\rhoi) =  \frac{4\pi\rvir^3\rhoz\fVz}{3+l}  \left(\frac{\rhoi}{\rhoz}\right)^{m+1}~.
\end{equation}
Similarly, the average column along the
sightline (eqn.~\ref{eq: NH Rimp avg single rho}) is generalized in a multi-density CGM to the average AMD, i.e.\ the average column per decade in density
\begin{equation}\label{eq: NH Rimp avg}
\langle\AMDfrac(\rhoi,\Rimp)\rangle = \nHi \int \fV(\rhoi,R)\deriv s ~.
\end{equation}
The covering factor of clouds with density $\rho$ is (in the limit that same-$\rho$ clouds do not overlap along a sightline) 
\begin{equation}
\label{eq: CF Rimp}
\CF(\rhoi,\Rimp) = \frac{3}{4\rc}\int \fV(\rhoi,R)\deriv s  ~,
\end{equation}
and from eqns.~(\ref{eq: NH Rimp avg})--(\ref{eq: CF Rimp}) we get 
\begin{equation}\label{eq: AMD vs. density structure}
 \rc(\rhoi) = \frac{3\langle\AMDfrac(\rhoi,\Rimp)\rangle}{4\nHi\CF(\rhoi,\Rimp)} ~,
\end{equation}
similar to the expression in the single density model (eqn.~\ref{eq: rc single rho}).
In the following section we discuss how $\langle\AMD\rangle$
and $\CF(\rhoi,\Rimp)$ can be constrained from absorption line measurements of ionic column densities.

\subsection{Absorption Features in a Multi-Density CGM}\label{sec: analytic}

What are the ionic columns expected in a multi-density CGM?
The column of ion $\ion$ along a given line of sight is equal to
\begin{equation}\label{eq: nion}
 \Nion = \frac{\el}{\rm H} \sum_{\rhoi} \AMDfrac(\rhoi)  \fion 
\end{equation}
where $\el/{\rm H}$ is the abundance of element $\el$ relative to hydrogen, assumed for simplicity to be independent of $\rho$, and $\fion$ is the fraction of $\el$ at ionization level $i$. We emphasize that eqn.~(\ref{eq: nion}) is accurate for a single line-of-sight, 
not only for an ensemble average. 
Now, in the absence of self-shielding $\fion$ is primarily a function of the ionization parameter $U$, defined as
\begin{equation}\label{eq: U}
 U\equiv \frac{\phii}{n_H c} = 2.0 \left(\frac{\rho}{\rhobar}\right)^{-1}\left(\frac{\phii}{\phiHM}\right) ~,
\end{equation}
where $\phii$ is the ionizing photon flux
incident on the cloud. 
The coefficient in eqn.~(\ref{eq: U}) originates from the choice of normalization, where we normalize $\rho$ by $\rhobar=7.2\times10^{-31} \g \cm^{-3}$,
the cosmic mean baryon density at $z=0.2$ (equivalently, $\nHbar = 3.2\times10^{-7} \cm^{-3}$), and we normalize $\phii$ by $\phiHM$, defined as 
\begin{equation}\label{eq: phiHM}
\frac{\phiHM}{c}=\frac{1}{c}\int_{\nu_0} 4\pi \frac{J_\nu}{h\nu} \deriv\nu = 6.4\times10^{-7} \cm^{-3} ~.
\end{equation}
In eqn.~(\ref{eq: phiHM}) $J_\nu$ is the UV background intensity at $z=0.2$ found by \citeauthor{HaardtMadau12} (2012, hereafter HM12), the integrand is from the Lyman edge ($\nu_0$) to infinity, and $c$ is the speed of light. 
The redshift is chosen to match the objects analyzed below. 

\begin{figure}
 \includegraphics{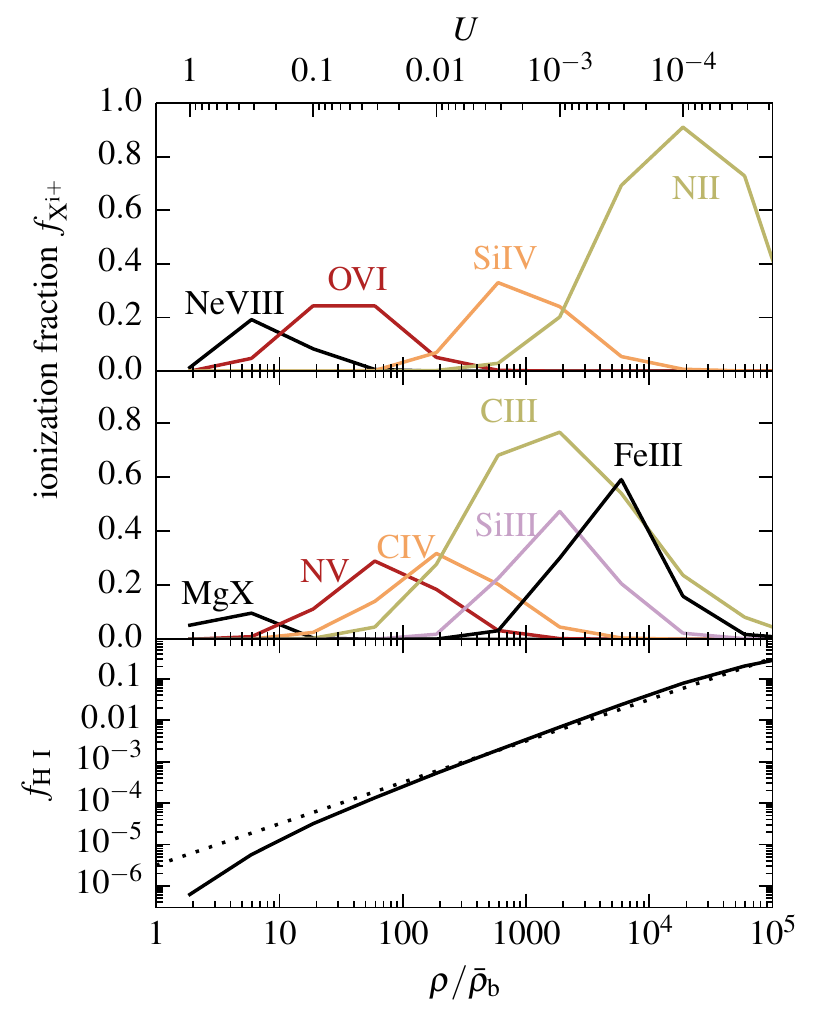}
\caption{
{\bf (Top panels)}
Ionization fractions of various ions versus gas density. Only ions with ionization energy $>1\ryd$ are shown. Calculations are done by \cloudy, assuming gas which is optically thin to hydrogen-ionizing photons ($\NHI\ll10^{17.2}\cm^{-2}$) and is illuminated by the UV-background at $z=0.2$ from HM12. 
The density is normalized by the baryonic cosmic mean density $\rhobar$ at $z=0.2$. 
The implied ionization parameter $U$ (eqn.~\ref{eq: U}) is noted on top. 
Note that each metal ion exists in significant quantities only within a dynamical range of $\sim10$ in $U$, which allows one to associate each metal ion with a specific gas density. 
{\bf (Bottom panel)}
The \hi-fraction under the same conditions as in the top panels. The dotted line is an analytic approximation used in eqn.~(\ref{eq: hi fraction}).
}
\label{fig: ionfractions}
\end{figure}

The dependence of $\fion$ on $U$ for several ions with ionization potential $>1\ryd$ is shown in Figure~\ref{fig: ionfractions}. 
The calculations are performed with version 13.03 of \cloudy\ (last described by \citealt{Ferland+13}) on a slab of gas illuminated by the UV background at $z=0.2$ calculated by HM12. The gas is assumed to have solar metallicity and a column of $\NH=10^{18.5}\cm^{-2}$, though these two parameters have a small effect on $\fion$ as long as the slab is optically thin at $1\ryd$. The effect of deviations from the HM12 UV background are addressed in the discussion. 
Figure~\ref{fig: ionfractions} demonstrates that each metal ion exists in significant quantities only within a dynamical range of $\sim10$ in $U$, so to first order we can keep only one term in the sum in eqn.~(\ref{eq: nion}) for each ion.
Assuming for example a CGM composed of three densities $\rho_0=50\rhobar$, $\rho_1=500\rhobar$, and $\rho_2=5000\rhobar$, we get
\begin{eqnarray}\label{eq: AMD purpose}
 \novi    &=& 1.2 \times 10^{-4} \frac{Z}{\zsun}\cdot\AMDfrac(\rho=50\rhobar) \nonumber \\
 \nsiiv   &=& 0.9 \times 10^{-5} \frac{Z}{\zsun}\cdot\AMDfrac(\rho=500\rhobar) \nonumber \\
 \nnii    &=& 4.2 \times 10^{-5} \frac{Z}{\zsun}\cdot\AMDfrac(\rho=5000\rhobar) 
\end{eqnarray}
where we replaced $\el/{\rm H}$ with the solar abundance of each element (\citealt{Asplund+09}) multiplied by the metallicity in solar units $Z/\zsun$. 
Equation~(\ref{eq: AMD purpose}) demonstrates that the AMD $\AMD$, which is the column density associated with a given density phase, can be directly constrained from observed ion columns given the metallicity (which one also typically fits for).

Eqn.~(\ref{eq: AMD purpose}) also suggests a close relation between the covering factors of the different ions and the covering factor of gas with different densities,
i.e.\ $\CF(50\rhobar,\Rimp)$ is roughly equal to the chance that a sightline with impact parameter $\Rimp$ shows \ovip\ absorption, $\CF(500\rhobar,\Rimp)$ is equal to the chance a sightline shows \Siivp\ absorption, and $\CF(5000\rhobar,\Rimp)$ is equal to the chance a sightline shows \niip\ absorption.

Thus, absorption line observations combined with photoionization modeling provide constraints on the two observable quantities $\CF(\rhoi,\Rimp)$ and $\AMD$, which can then be used to determine the parameters of the multi-density cloud model.

\subsection{A Hierarchical Cool CGM}\label{sec: hierarchical}

\begin{figure}
\includegraphics{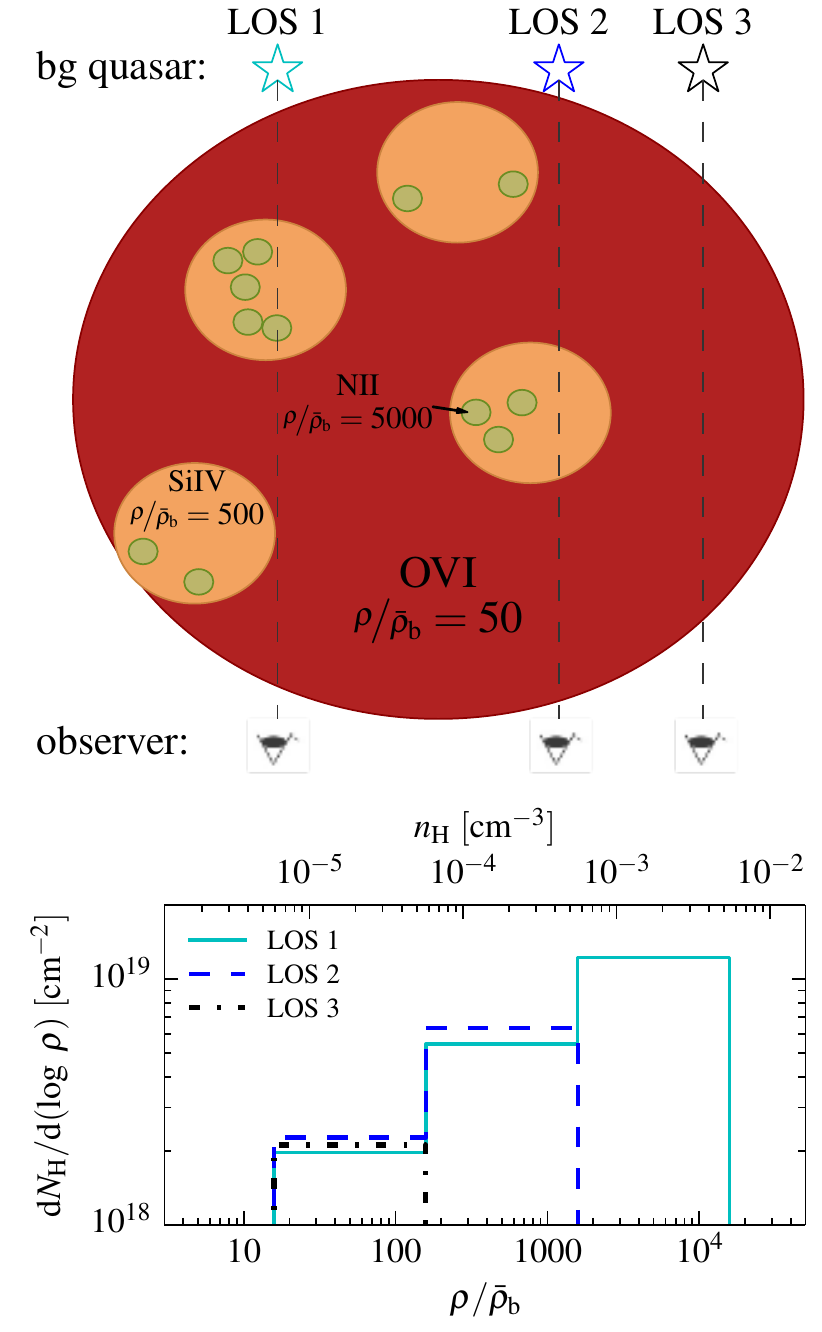}
\caption{
{\bf (Top)}
A schematic illustration of a hierarchical CGM cloud, where small high-density clouds are embedded in larger low-density clouds. The entire cloud is assumed to be photoionized by the UV background. 
For each phase we mark the assumed gas density and a characteristic ion.
The free parameters of this assumed cloud structure are the size of clouds at each density $\rc(\rhoi)$ and the filling factor of clouds at each density and distance from the galaxy $\fV(\rhoi,R)$. 
Also plotted are three different lines of sight, where LOS~1 intersects gas with all the considered densities, LOS~2 intersects the medium- and low-density gas, and LOS~3 only the low-density gas. 
{\bf (Bottom)}
The Absorption Measure Distribution (AMD, gas column per decade in density) of the three LOSs, for an assumed \ovip-cloud size of $30\kpc$. 
The AMD of LOS~2 is roughly equal to the AMD of LOS~1 truncated at $\rhomax\approx2000\rhobar$.
Similarly, the AMD of LOS~3 is roughly equal to the AMD of LOS~1 truncated at $\rhomax\approx200\rhobar$.
We treat COS-Halos sightlines as different LOSs through the same CGM, 
and hence COS-Halos sightlines are expected to have similar AMDs which differ only in the maximum probed density $\rhomax$. 
}
\label{fig: schema}
\end{figure}

The kinematic alignment of low-ions and high-ions mentioned in the introduction suggests a spatial correlation between the high-density and low-density clouds. 
Hence, we make an additional assumption on the structure of the multi-density cool CGM, 
that small high-density clouds are hierarchically embedded in larger low-density clouds.
Using the example from the previous section 
we imagine an idealized case of three gas phases, one tracing \ovip, one tracing \Siivp, and one tracing \niip, as pictured in the top panel of Figure~\ref{fig: schema}.

Figure~\ref{fig: schema} shows three possible lines-of-sight (LOS) through the hierarchical cloud, where LOS~1 intersects gas with all the considered densities, LOS~2 intersects the medium- and low-density gas, and LOS~3 intersects only the low-density gas. 
We hence expect all three LOSs to exhibit $\ovip$ absorption, but only LOS~1 and LOS~2 should exhibit appreciable \Siivp\ absorption, and only LOS~1 to exhibit appreciable \niip.
This property of the hierarchical structure that low-ions are always expected to be associated with high-ions, but not vice-versa, is consistent with absorption line observations (see below), and hence supports our assumption of a hierarchy.

The bottom panel of Figure~\ref{fig: schema} shows the AMDs of the three LOSs pictured in the top panel. The AMDs are calculated from the pathlength of each LOS through each phase, as shown in the Figure, where the assumed scale is such that the \ovip-cloud size is $\rz=30\kpc$. 
Note that the AMD of LOS~2 is roughly equal to the AMD of LOS~1, truncated at $\rhomax\approx2000\rhobar$.
Similarly, the AMD of LOS~3 is roughly equal to the AMD of LOS~1 truncated at $\rhomax\approx200\rhobar$. 
That is, the AMDs of different LOSs can be considered as segments of some `universal' AMD which spans all $\rho$,
but which are truncated at some maximum density $\rhomax$ which is specific to the individual LOS.
In other words, if the AMD of individual sightlines is a power-law of the form
\begin{equation}\label{eq: AMD}
 \AMDfrac = \dNz \left(\frac{\rho}{\rhoz}\right)^{\beta} ~~\{\rhoz < \rho < \rhomax\} ~,
\end{equation}
then all sightlines have roughly the same $\dNz$ and $\beta$, while they differ significantly only in their $\rhomax$. 
Except the truncation, the differences between the three AMDs are small, of order unity. 
These small differences originate from the different possible pathlengths of a LOS through a given cloud, 
and from the different number of same-$\rho$ clouds intersected by a LOS.

Our approach to fitting photoionization models to CGM absorption line data is thus as follows. 
We utilize the standard approach of galaxy-selected samples of CGM absorbers (such as COS-Halos) to treat individual sightlines through the CGM
of similar galaxies as different sightlines through the same CGM. 
We fit the same $\dNz$ and $\beta$ to all objects, while the difference between the absorption features in different objects is set by the maximum density $\rhomax$ encountered along the sightline. We also fit an individual gas metallicity $Z$ to each object, which is held constant across all $\rho$ phases\footnote{The metallicity may in principle depend on gas density, for example in the picture suggested by \cite{Schaye+07}. We defer exploring this possibility to future work.}. That is, we model the observed absorption features of all objects in the sample {\it simultaneously}, using two universal parameters ($\dNz$ and $\beta$) plus two parameters per object ($\rhomax$ and $Z$).
Our approach hence has an advantage over previous studies of galaxy-selected samples of CGM absorbers. While the conventional approach is to model each sightline separately, and then discuss the aggregate CGM properties via some average over these noisy individual absorption models, our approach combines the constraints from all objects during the absorption line modeling itself
to obtain one high S/N fit for the model parameters. 

Before elaborating on the numerical absorption line modeling in the next section, we provide some physical intuition for the parameters $\dNz$ and $\beta$ introduced in eqn.~(\ref{eq: AMD}). 
In the hierarchical model $\CF(\rho,\Rimp)$ is equal to the fraction of sightlines which have $\rhomax>\rho$:
\begin{equation}\label{eq: CF def}
 \CF(\rho,\Rimp) = P(\rhomax>\rho ~| \Rimp) ~,
\end{equation}
because the hierarchical assumption implies that a sightline with truncation density $\rhomax$ intersects all lower density phases.
Combining eqn.~(\ref{eq: CF def}) with eqn.~(\ref{eq: AMD}) implies that the average AMD of an ensemble of sightlines is equal to 
\begin{equation}\label{eq: AMD av}
 \langle\AMDfrac(\rhoi,\Rimp)\rangle = \CF(\rhoi,\Rimp) \cdot \dNz \left(\frac{\rhoi}{\rhoz}\right)^{\beta} ~.
\end{equation}
Now, by comparing eqn.~(\ref{eq: AMD av}) with eqn.~(\ref{eq: AMD vs. density structure}) we get \begin{equation}\label{eq: AMD vs. density structure values1}
 \dNz\left(\frac{\rhoi}{\rhoz}\right)^\beta = \frac{4}{3}\rc(\rhoi)\nHi ~.
\end{equation}
Since $(4/3)\rc$ is the average pathlength through a cloud, eqn.~(\ref{eq: AMD vs. density structure values1}) implies that $\dNz (\rhoi/\rhoz)^{\beta}$ equals the average column of a sightline through a single cloud with density $\rhoi$. 
Also, since $\rc=(\rhoi/\rhoz)^\alpha$ (eqn.~\ref{eq: density structure}) we get
\begin{equation}\label{eq: AMD vs. density structure values2}
\rz = \frac{3\dNz}{4n_{\rm H,0}}~, ~~~\alpha=\beta-1 ~.
\end{equation}
These last two relations are accurate in the no-overlap limit, in which eqn.~(\ref{eq: AMD vs. density structure}) is derived. 
When allowing for overlaps of same-$\rho$ clouds along the line of sight, $\dNz (\rhoi/\rhoz)^{\beta}$ equals the average column in sightlines which intersect {\it at least} one cloud with density $\rhoi$. 
So, eqn.~(\ref{eq: AMD vs. density structure values2}) overestimates $\rc(\rhoi)$ by a factor equal to the mean number of $\rhoi$-clouds along such sightlines, which can be estimated numerically.

\subsection{Numerical Photoionization Modeling}\label{sec: numerical}

To derive a more accurate solution of the expected absorption features as a function of $\dNz$, $\beta$, $\rhomax$, and $Z$, we use the photoionization code \cloudy. Specifically, \cloudy\ allows us to drop the crude approximation in eqn.~(\ref{eq: AMD purpose}) that each ion originates exclusively from a single phase, and also to account for the effects of self-shielding.

\cloudy\ can calculate the ionization structure of a slab where the gas density varies as a function of the H-column measured from the slab surface $\NH'$ (note that $\NH'$ is a coordinate within the slab, in contrast with $\NH$ which is the total column observed along some line of sight). 
To find the dependence of $\rho$ on $\NH'$ which reproduces a desired AMD, we integrate eqn.~(\ref{eq: AMD}) ($\beta\neq0$):
\begin{equation}\label{eq: NH'}
 \NH'(\rho) =  \frac{\dNz}{\beta\ln 10}\left[ \left(\frac{\rho}{\rhoz}\right)^{\beta} - 1\right] ~,
\end{equation}
which implies 
\begin{equation}\label{eq: for cloudy}
 \rho(\NH') = \rhoz \left[ 1+\frac{\NH'}{\dNz/(\beta\ln 10)}\right]^{\frac{1}{\beta}} ~.
\end{equation}
Equivalently, for $\beta=0$ we get $\NH'(\rho) = \dNz \log(\rho/\rhoz)$ and $\rho(\NH')=10^{\NH'/\dNz}\rhoz$.
In all calculations we set $\rhoz=20\rhobar$, or equivalently $\nz=0.6\times 10^{-5}\cm^{-2}$, since layers with $\rho<20\rhobar$ are so highly ionized that they do not change the predicted columns of ions observed in the COS-Halos sample analyzed below. Observations of higher ionization lines such as $\neviiip~\lambda\lambda770,~780$ are required to constrain the properties of these low density 
layers (see Figure~\ref{fig: ionfractions}). Available \neviiip\ observations are addressed in the discussion.
Assuming higher values for $\rhoz$ yields poorer fits.

In total, we run 168 \cloudy\ models, with $\beta = 0.01, 0.125,  0.25,  0.375,  0.5,  0.625,$ or 0.75; $\log\ \dNz = 17.5, 17.8,  18.1,  18.4,  18.7$, or $19$; and $Z/\zsun= 0.1,0.3,1$ or 3. For each combination of $\dNz$ and $\beta$, we run \cloudy\ where $\rho$ within the slab varies according to eqn.~(\ref{eq: for cloudy}).
We assume a HM12$(z=0.2)$ incident spectrum and solar relative abundances in all calculations. 
The stopping criterion of the models is set to an arbitrary large $\nhi$ of $10^{20}\cm^{-2}$.
In order to reproduce the truncation of the AMD at $\rho=\rhomax$ as discussed
above, for each layer in the slab calculated by \cloudy\ we record the gas density and the column of each ion from the illuminated surface up to this layer. 
Thus, for each \cloudy\ model with parameters $(\beta, \dNz, Z)$ and for each ion $\ion$, we get a predicted column $\Nexp$ as a function of $\rhomax$, yielding a four-dimensional grid for each ion $\Nexp({\ion},\beta, \dNz, Z, \rhomax)$.
To derive $\Nexp$ for parameters which are not in the grid, we interpolate between the nearest grid values. 
$\hi$ is treated as any other ion in the fit.

For a given sample of $k$ quasar-galaxy pairs,
we use the Levenberg-Marquardt  
algorithm (\citealt{Press+92}) to find the best-fit for all ion columns with $2k+2$ free parameters: $Z$ and $\rhomax$ for each object, plus one universal $\beta$ and one universal $\dNz$.  
For comparison, the standard photoionization modeling approach where a constant-density model is fit to each absorber
has $3k$ parameters, namely $\rho, \NH$, and $Z$ for each object ($\rho$ is often replaced with $U$).
Similarly, an absorber-specific model which fits two densities per object has $5k$ parameters (two $\rho$, two $\NH$, and one $Z$ per object). 

The likelihood $\mathscr{L}$ to be maximized is 
\begin{eqnarray}\label{eq: likelihood}
\ln & &\mathscr{L}(\beta, \dNz, \{Z\}, \{\rho_{{\rm max}}\}| \{ \Nobs\}) = \nonumber \\
         & & -\frac{1}{2}\cdot\frac{1}{0.2^2} \sum_j  \sum_{\ion} \left( \log \frac{\Nexp({\ion},\beta, \dNz, Z_j, \rho_{{\rm max}; j})}{\Nobs(j,\ion)} \right)^2 \nonumber\\
\end{eqnarray}
where $j=1..k$ is an index over all objects in the sample,
$\Nobs(j, \ion)$ is the observed column of ion $\ion$ in object $j$,
and curly brackets denote a set of parameters or measurements. This likelihood is similar to the likelihoods in \cite{Crighton+15} and \cite{Fumagalli+16}. 
We assume a common error of $0.2\dex$ on all measurements, larger than the typical measurement uncertainties of $\lesssim 0.1\dex$. This error accommodates expected variations in $\AMD$ of order 50\% between different sightlines, as seen in the lower panel of Figure~\ref{fig: schema} and further justified below. This error also accommodates for relative abundance deviations from Solar. 
Note that since a constant error is assumed on all column measurements, the absolute value of the error does not affect the result of the best-fit.
 
In cases where a measurement $\Nobs(j, \ion)$ is an upper or lower limit, we adopt an approach similar to \cite{Crighton+15} where we assume a one-sided Gaussian with error $0.2\dex$ beyond the limit. If the predicted column conforms to the limit then the argument of the sum
in eqn.~(\ref{eq: likelihood}) is assumed to be zero. 
Since $-2\ln\mathscr{L}$ is basically a chi-square except for how we treat limits, for ease of notation we henceforth use $\chi^2\equiv-2\ln \mathscr{L}$.

\subsection{The relation between $N_{\rm HI}$ and $\rhomax$}\label{sec: nhi}

Another noteworthy property of the hierarchical model is the close relation between $\nhi$ and $\rhomax$. This relation can be derived by noting that the $\hi$ fraction can be approximated as (bottom panel of Figure~\ref{fig: ionfractions})
\begin{equation}\label{eq: hi fraction}
 f_\hi(\rho) \approx \left(\frac{U}{6\times10^{-6}}\right)^{-1} = 3\times 10^{-6} \left(\frac{\rho}{\rhobar}\right) ~,
\end{equation}
where in the second equality we assume $\phii=\phiHM$. 
Using eqn.~(\ref{eq: hi fraction}) in eqn.~(\ref{eq: AMD}) we get that the contribution to $\NHI$ from each phase is
\begin{equation}\label{eq: nhi}
 \frac{\deriv\nhi}{\deriv\log\rho}(\rhoi) = f_\hi(\rhoi) \AMDfrac \propto \rhoi^{\beta+1} ~.
\end{equation}
Eqn.~(\ref{eq: nhi}) implies that for $\beta\gtrsim0$ the value of $\nhi$ is determined by the gas with the highest density along the line of sight, i.e.
\begin{equation}
 \NHI \sim f_\hi(\rhomax)\AMDfrac(\rhomax) ~.
\end{equation}

If we assume $\beta=0$ and $\dNz=10^{18.5}\cm^{-2}$, which are close to the best-fit values found below, we get
\begin{eqnarray}\label{eq: nhi by rhomax values}
 \NHI &\sim& f_\hi(\rhomax)\AMDfrac(\rhomax) = f_\hi(\rhomax)\dNz\left(\frac{\rhomax}{\rhobar}\right)^\beta \nonumber\\
      &\approx&  0.9\times 10^{13} \frac{\rhomax}{\rhobar} \left(\frac{\dNz}{10^{18.5}\cm^{-2}}\right) \cm^{-2} ~.
\end{eqnarray}
where we replaced $f_\hi(\rhomax)$ with the expression in eqn.~(\ref{eq: hi fraction}). 
Eqn.~(\ref{eq: nhi by rhomax values}) suggests that the large dynamical range of $>10^6$ in $\NHI$ observed in the COS-Halos sample (\citealt{Tumlinson+13}) is due to a similarly large dynamical range in the maximum gas density encountered along the sightline $\rhomax$.

\section{Application to the COS-Halos sample}\label{sec: application}

\subsection{The COS-Halos sample}

We use data from the COS-Halos survey (\citealt{Tumlinson+11,Tumlinson+13,Thom+12,Werk+12,Werk+13,Werk+14}) 
of CGM gas surrounding 44 galaxies at low-redshift ($z\sim0.2$) with luminosity $0.1\,L^* < L < 3\,L^*$. 
COS-Halos observed 39 UV-bright quasars within an impact parameter $\Rimp < 160\kpc$ from the sample galaxies
using the {\it Cosmic Origins Spectrograph} (COS; \citealt{Green+12}) on board the {\it Hubble Space Telescope} (HST).
\cite{Tumlinson+13} discuss the design and execution of the survey. 
For our purposes, we use 580 of the 589 metal ion column measurements listed in \cite{Werk+13}, including detections and limits of all ions up to \ovip, and excluding only the nine measurements which are marked as `blended and saturated'. These column measurements were derived using the apparent optical depth method (\citealt{SavageSembach91}) on a velocity range typically $-200 < v <200\kms$ from the galaxy redshift. The 379 non-detections (65\%) are given as 2$\sigma$ upper limits, while another 74 absorption features (13\%) are saturated, so they are treated as lower limits.
In cases where there are multiple detected transitions for a certain ion in a given object, we use the weighted mean ion column listed in Table 3 in \cite{Werk+13}. 

We supplement the metal ion columns with the 44 \hi-column measurements listed in Table~1
of \cite{Werk+14}. Four objects have no detection of a Lyman absorption feature, so their $\NHI$ are treated as upper limits. In another 22 objects the allowed range of $\NHI$ is an order of magnitude or more since the Lyman features are saturated. We treat these measurements as two-sided limits, where the contribution to $\ln\mathscr{L}$ (eqn.~\ref{eq: likelihood}) is zero if the predicted $\NHI$ falls within the allowed range, and an error of $0.2\dex$ is assumed beyond the allowed range. 
Including these $\NHI$ measurements, we fit a total of 624 detections and limits. 

\subsection{Fit results}

\begin{figure}
\includegraphics{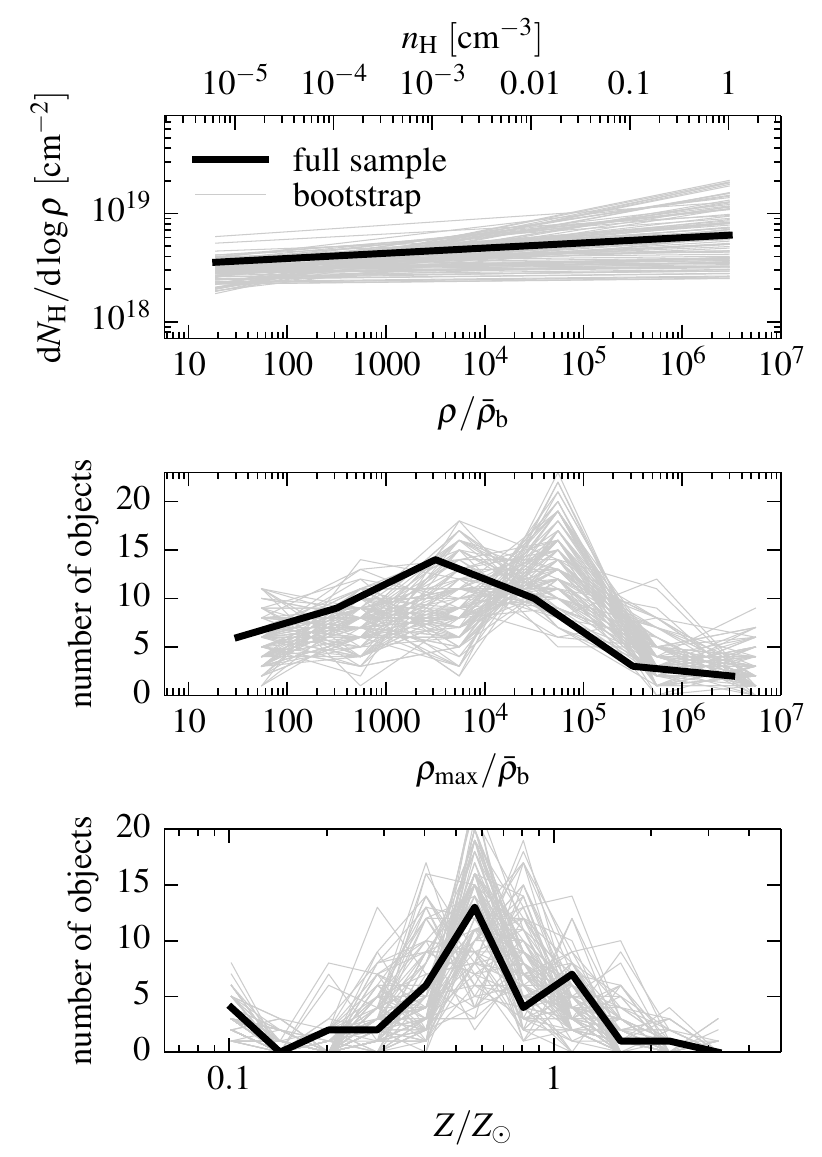}
\caption{
Best-fit parameters to the 44 objects in the COS-Halos sample. Black lines are the best-fit to the entire sample, 
while gray lines are best-fits to different choices of 44 objects with replacement, which provide a bootstrap estimate of the error in our procedure. 
{\bf (Top)}
The best-fit normalization and slope of the AMD, which are common to all objects. 
Note the AMD slope is flat, with a constant characteristic column  per decade in density of $\AMD\sim10^{18.5}\cm^{-2}$.
{\bf (Middle)}
The distribution of the 44 best-fit maximum densities $\rhomax$. 
In a given object, the observed AMD is the power law seen in the top panel truncated at the fit $\rhomax$ (eqn.~\ref{eq: AMD}). Note that $\rhomax$ span a large dynamical range of $\approx10^5$.
{\bf (Bottom)}
The distribution of the 44 best-fit metallicities in the COS-Halos sample. 
}
\label{fig: AMD}
\end{figure}

\begin{figure*}
\includegraphics{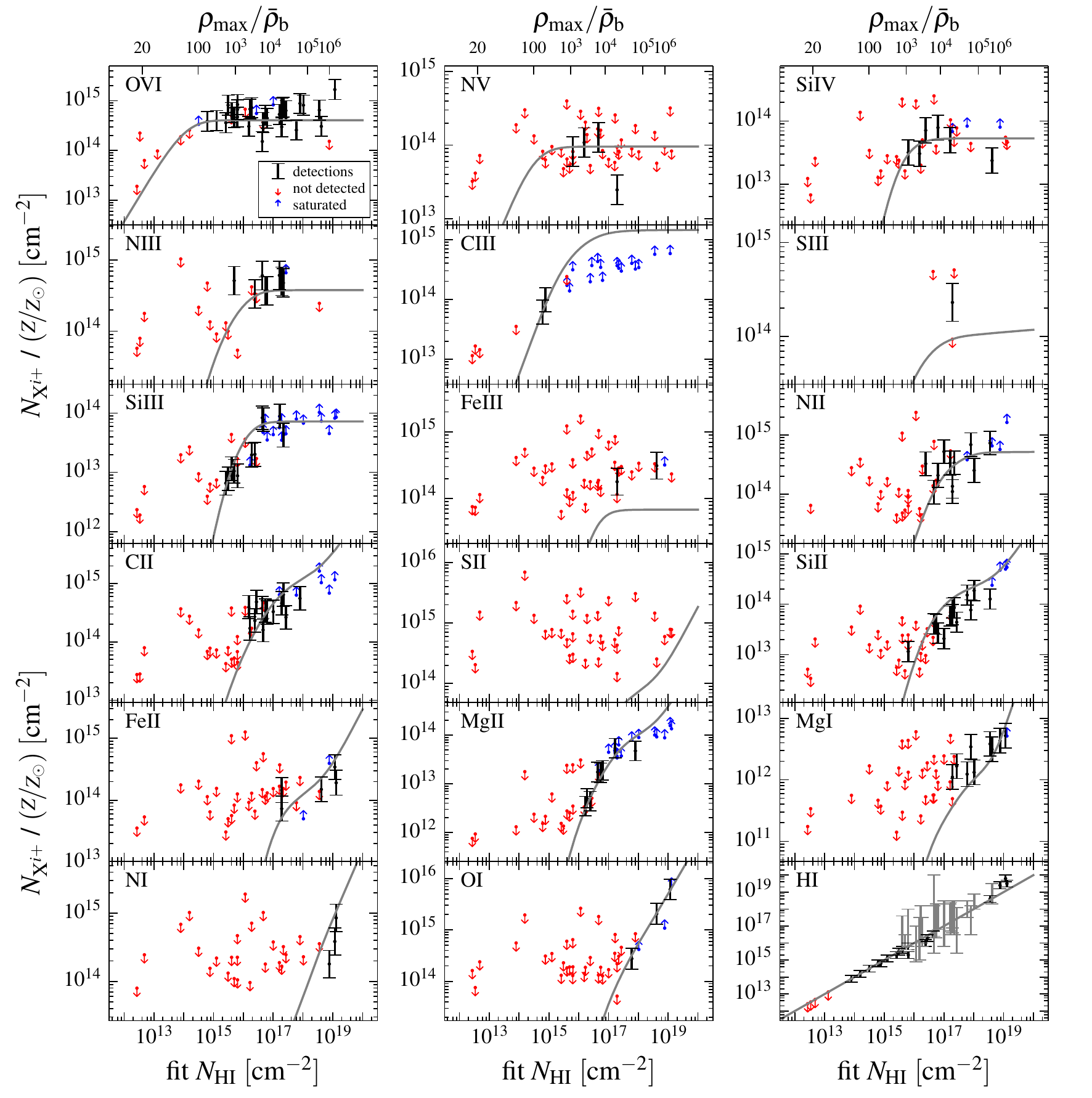}
\caption{
Comparison of observed ion columns in the COS-Halos sample with the ion columns implied by the best-fit model (Figure~\ref{fig: AMD}). 
Each panel shows a specific ion in all COS-Halos objects, where the panels are ordered by decreasing ionization energy. 
The upper horizontal axis of each panel is the value of $\rhomax$ fit to each object,
which is strongly related to the fit $\NHI$ in the lower horizontal axis (see \S\ref{sec: nhi}). 
Gray lines plot the fit ion columns vs.\ the fit $\NHI$ assuming $Z=\zsun$.
Observed ion columns are marked by either black error bars with the assumed uncertainty of 0.2\dex, or blue (red) arrows for lower (upper) limits.
Measurements of $\NHI$ in the lower-right panel with a large possible range are plotted as gray error bars. 
The observed metal columns are normalized by the metallicity in solar units fit to each object (typically $\sim0.6\zsun$). 
The generally good agreement between the expected and observed ion columns supports our assumed hierarchical density structure. 
A quantitative assessment of the goodness-of-fit is done in Figure~\ref{fig: cross validation} below.
}
\label{fig: fit by ion}
\end{figure*}

We apply the fitting algorithm described in \S\ref{sec: numerical} to the 624 column measurements. 
The best-fit AMD normalization $\dNz$ and AMD slope $\beta$ are shown in the top panel of Figure~\ref{fig: AMD}. 
To estimate the error on $\dNz$ and $\beta$ we use the bootstrap method, where we choose with replacement $k=44$ objects from the COS-Halos sample and run the fit on this new sample, repeating this process 100 times. 
The found AMD is 
\begin{equation}
 \AMDfrac = \dNz \left(\frac{\rho}{20\rhobar}\right)^{\beta} ~~\{20\rhobar < \rho < \rhomax\} \nonumber 
\end{equation}
with 
\begin{equation}\label{eq: AMD fit}
\beta =  0.05 \pm 0.05 ~,~~ 
\dNz  = 10^{18.54 \pm 0.1}\cm^{-2}   ~,
\end{equation}
and the distribution of $\rhomax$ shown in the middle panel of Figure~\ref{fig: AMD}. The errors on $\beta$ and $\dNz$ quoted in eqn.~(\ref{eq: AMD fit}) are marginalized errors, since these two parameters are covariant.
Note the fit values of $\rhomax$ span a large dynamical range of $\approx10^5$. 
The implications of this $\rhomax$ distribution are discussed in the next section.  

The fit $Z$ distribution is shown in the bottom panel of Figure~\ref{fig: AMD}. 
The typical $Z$ in the CGM of COS-Halos galaxies is found to be $Z \approx 0.6\zsun$, with $68\%$ of the objects in the range of $0.3 < Z < 1.1 \zsun$. The plotted $Z$ distribution excludes the four objects in which all ion measurements are upper limits, and therefore $Z$ cannot be constrained. Four additional objects have only \hi\ detections, and hence the fit $Z$ of 0.1, 0.1, 0.1, and 0.3$\,\zsun$ in these objects are upper limits. 

Figure~\ref{fig: fit by ion} compares the ion columns calculated by our best-fit model to the observed columns. 
Each of the 18 panels shows a specific ion, where the panels are ordered by decreasing ionization energy. 
The gray lines are the calculated ion columns as a function of $\rhomax$ (noted on top), assuming $Z=\zsun$. 
The fit $\nhi$, which is closely related to $\rhomax$ (see approximation in eqn.~\ref{eq: nhi by rhomax values}), is noted at the bottom of the Figure.
The observed ion columns are plotted versus the
best-fit $\rhomax$ (and also $\NHI$)
of the relevant object, and the metal ion columns are normalized by the fit $Z/\zsun$.
Detections are marked by error bars with our assumed uncertainty of 0.2\dex\ (eqn.~\ref{eq: likelihood}), while upper and lower limits are marked by colored arrows. 
Therefore, each panel in Figure~\ref{fig: fit by ion} has up to 44 data points, one from each foreground galaxy in the sample.

By eye, the expected and observed ion columns are generally in good agreement. The best-fit yields a maximum likelihood estimate (eqn.~\ref{eq: likelihood}) of $\chi^2\equiv -2\ln\mathscr{L} = 371$, for a sample of data points of which 148 are detections and 476 are limits, using 90 free parameters (a universal $\dNz$ and $\beta$, and $\rhomax$ and $Z$ for each of the $k=44$ objects). However, since a large fraction of data points are limits, the model is non-linear and the usual $\chi^2$ per degree of freedom intuition does not hold (\citealt{Andrae+10}). Therefore, in the next section we perform an alternative assessment of the goodness-of-fit using the cross-validation technique.

\begin{figure*}
\includegraphics{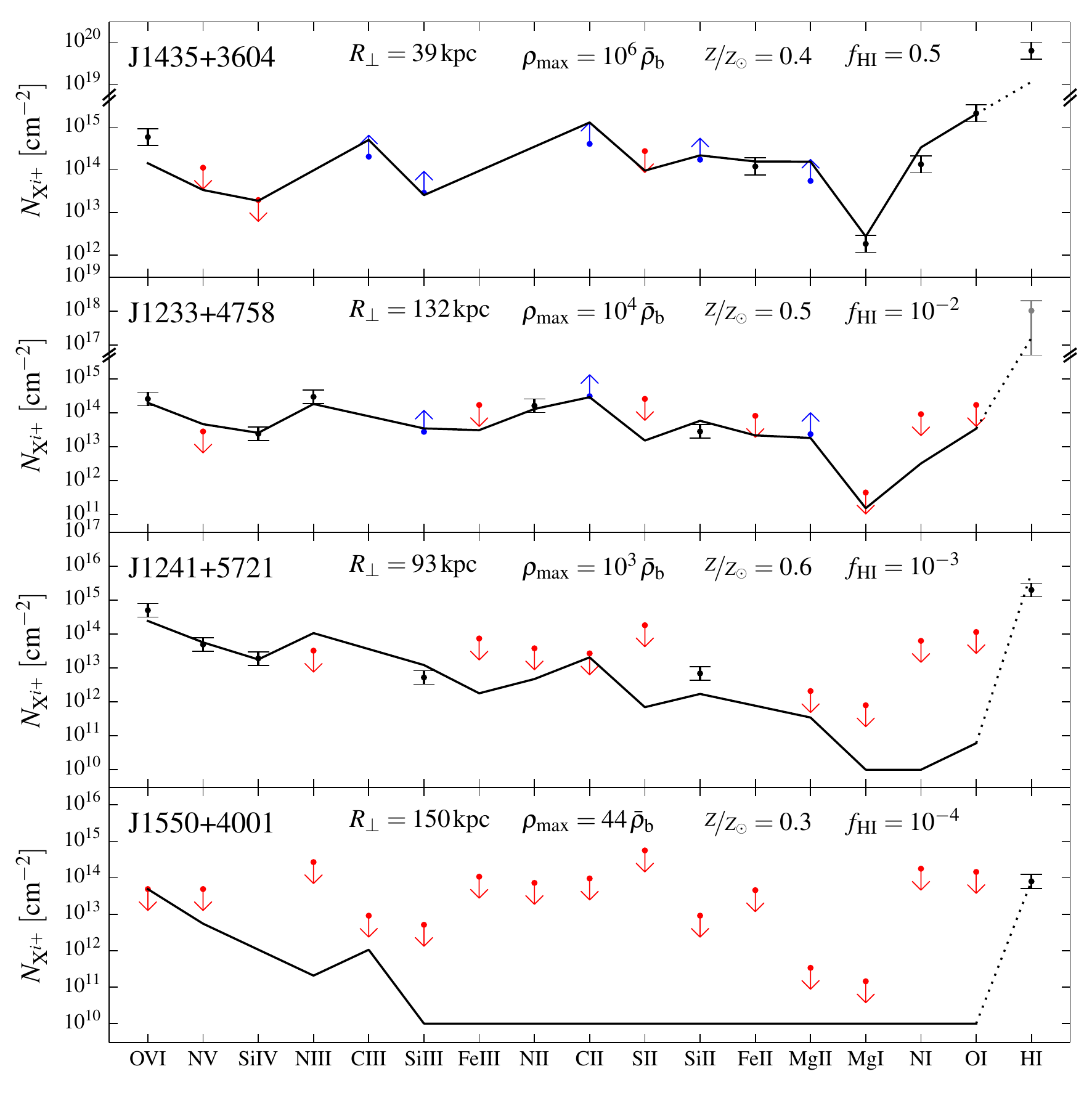}
\caption{
Comparison of observed ion columns in four COS-Halos objects with the ion columns implied by the best-fit model (Figure~\ref{fig: AMD}). 
Objects are selected to span the huge dynamical range of $>10^6$ in $\nhi$ seen in the COS-Halos sample. 
Fits to the remaining objects are available online. 
The solid lines denote the fit metal ion columns, while the fit $\nhi$ is connected by a dotted line. 
Observations are marked as in Figure~\ref{fig: fit by ion}. 
For each object we note the impact parameter $\Rimp$, the best-fit $\rhomax$ and best-fit $Z$, and the \hi-fraction $f_\hi$. 
The generally good agreement between the expected and observed ion columns supports our assumed hierarchical density structure. 
A quantitative assessment of the goodness-of-fit is done in Figure~\ref{fig: cross validation} below.
}
\label{fig: fit by obj}
\end{figure*}

The $\ovip$ panel in Figure~\ref{fig: fit by ion} demonstrates that the predicted $\novi$ increases with increasing $\nhi$ only at $\nhi<10^{14.5}\cm^{-2}$ ($\rhomax<100\rhobar$),
and is independent of $\nhi$ at larger values of $\nhi$. 
This predicted independence of $\novi$ on $\nhi$ at large $\nhi$ is apparent in the COS-Halos observations, 
and has also been observed in other samples of intervening absorbers at low $z$ (\citealt{Danforth+14}), and also at $z\gtrsim 2$ (\citealt{Muzahid+12, Lehner+14}).
In the context of our model, this independence of $\novi$ on $\nhi$ occurs since \ovip\ is produced only in the low density layer with $U\sim 0.04$ ($\rho\sim 50\rhobar$), 
which the AMD in the upper panel of 
Figure~\ref{fig: AMD} shows has a characteristic column of $[\deriv\NH/\deriv\log\nH](\rho=50\rhobar) \sim 10^{18.5}\cm^{-2}$. 
According to eqn.~(\ref{eq: AMD purpose}), this characteristic H-column implies a characteristic $\novi \sim 10^{14.7}\cm^{-2}$ (for $Z=\zsun$), 
and a characteristic $\nhi\sim 10^{14.5}\cm^{-2}$ (eqn.~\ref{eq: hi fraction}). 
The observed $\nhi$\ will surpass this value of $10^{14.5}\cm^{-2}$ only if the line of sight passes through higher density layers, which have a lower ionization state and hence a larger \hi\ fraction. These layers will not however increase the observed $\novi$, since oxygen is less than five times ionized within them. 
Hence,
since lines-of-sight
which traverse the high-density layers also cross the low density \ovip-layer, 
then for all $\nhi\gg10^{14.5}\cm^{-2}$ the expected $\novi$ column is $\sim10^{14.7}(Z/\zsun)\cm^{-2}$. 
Thus, the independence of $\novi$ on $\nhi$ above some `transition-$\NHI$' is a direct result of our assumption that all the \ovip\ is photoionized and that the density structure of the absorbing gas is hierarchical. 
The value of this transition-$\NHI$ of $10^{14.5}\cm^{-2}$ is set by the typical $\NHI/\novi\sim 1$ expected in solar metallicity photoionized gas with $U\sim0.03$ where \ovip\ is most efficiently produced.  

The trend of increasing $\novi$ at small $\nhi$ and independence of $\novi$ on $\NHI$ 
at large $\nhi$, is also apparent in other ions in Figure~\ref{fig: fit by ion}, where the `transition-$\nhi$' increases with decreasing ionization energy. 
This characteristic behavior has also recently been observed in \civp-absorbers at $z\sim2$ (\citealt{Kim+16}). 
The increase in the transition-$\nhi$ with decreasing ionization energy occurs since low-ions reside in high-density layers, which produce a larger
$\NHI$. Thus, ions with different ionization energies are created within different parts of the absorber. 
At $\nhi>10^{17.2}\cm^2$, the optical depth to \hi-ionizing radiation becomes substantial, and the ionization state of the gas is dominated by the optical depth rather than by the gas density. In these self-shielded regions of the cloud only the column of atoms and ions which are created by photons with energy $<1\ryd$ (e.g.\ \mgiip) increase with $\nhi$.

Figure~\ref{fig: fit by obj} compares the calculated and observed ions grouped by object, for a few COS-halos galaxies.
From the 40 objects with detections in at least one ion, we show the objects with the highest, 14\th-highest, 27\th-highest, and lowest $\rhomax$, in order to span the entire $\rhomax$ and $\nhi$ range.
The fits to the rest of the objects are available online\footnote{http://www2.mpia-hd.mpg.de/homes/stern/UniversalFits/}.
For reference, we note for each object the impact parameter $\Rimp$, the best-fit $Z$ and $\rhomax$, and the calculated $f_\hi=\nhi/\NH$.
As implied by Figure~\ref{fig: fit by ion}, Figure~\ref{fig: fit by obj} shows that the calculated ion columns are generally consistent with the observations. 
Figure~\ref{fig: fit by obj} also demonstrates that as $\NHI$ and $\rhomax$ increase (bottom panel to top panel) lower-ionization ions become more prominent.

The flat AMD slope we find (eqn.~\ref{eq: AMD fit}) implies that the total gas column $\NH$ scales only logarithmically with $\rhomax$ (eqn.~\ref{eq: NH'}). Thus, in the 40 objects with at least one detected absorption feature we find a small dispersion in total column of $\NH = 10^{18.9}\pm 0.3\dex$, despite the large range in $\rhomax$.
This result is consistent with the result of \cite{Prochaska+04}, who deduced $\NH = 10^{18.7}\pm 0.3\dex$ in five of six absorbers along the line of sight to PKS~0405-123. 

\subsection{Comparison with absorber-specific modeling}

\begin{figure}
\includegraphics{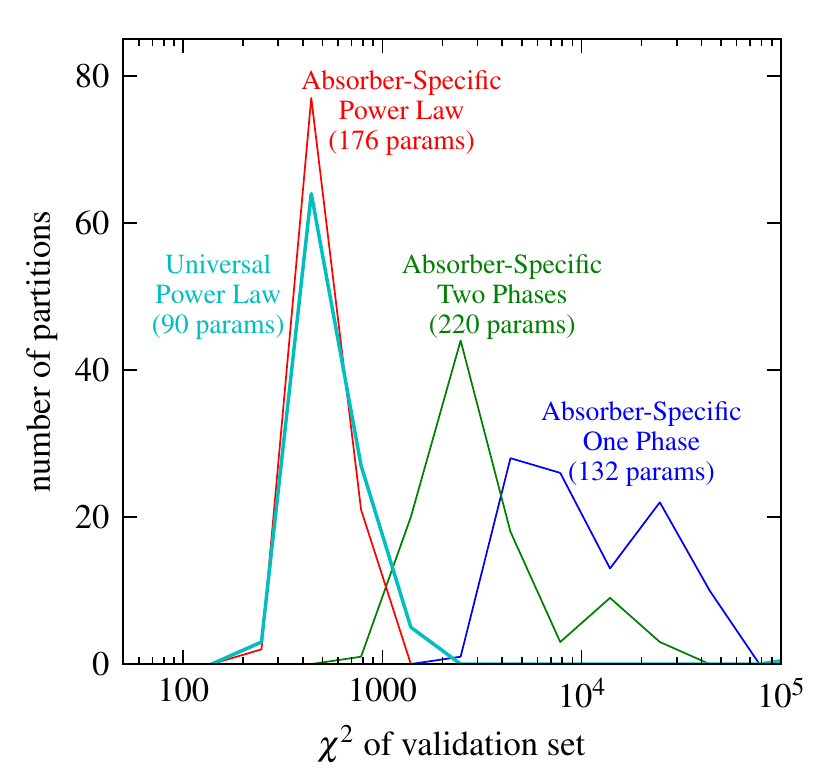}
\caption{
Comparison of the universal fit applied in this work with absorber-specific fits, using the cross-validation technique. 
The total number of free parameters in each model is noted. 
The universal model assumes a power-law density distribution which is common to all absorbers. 
The absorber-specific models fit each of the 44 COS-Halos objects with either a single gas density per object, two gas densities per object, or a power-law density distribution per object. 
The lower $\chi^2$ found in the validation sets of the power-law models suggest they have superior predictive power over the single- and double-density models 
conventionally used in the literature. 
}
\label{fig: cross validation}
\end{figure}
As can be seen in Figures~\ref{fig: fit by ion}--\ref{fig: fit by obj} and discussed above, the quality of the fit is reasonable, and explains the general trends in the data, supporting our suggested CGM density structure. However, a quantitative estimate of the goodness-of-fit is not straightforward, since the found $\chi^2=371$ depends strongly on our assumed error of $0.2\dex$, which is somewhat uncertain, and the models are non-linear in the parameters such that the usual $\chi^2$ per degree of freedom intuition does not hold.
Also, we wish to compare the success of our universal model with the absorber-specific models typically used in the literature.
For this purpose it is not trivial to use a $\chi^2$-like goodness of fit criteria
since universal and absorber-specific models differ significantly in the number of free parameters.
Therefore, in this section we use the {\it cross-validation} technique to compare the predictive power of our universal model with the predictive power of absorber-specific models.
In cross-validation, one randomly partitions the data points into a `training set', on which the free parameters of the model are optimized, and then calculates the measure-of-fit (here the $\chi^2$ score) on the remaining `validation set' using the parameters optimized to fit the training set.
This procedure tests the ability of the model to predict unknown data points. 
The process is repeated several times with different partitions, which yields a distribution of $\chi^2$ per model. The resulting distributions of $\chi^2$ of the different models can then be compared to check which model has the strongest predictive power.

We compare our universal model with the following three absorber-specific models.
The first model is a constant-density absorber with a single gas density per line of sight. In this model, we find the values of $\nH$, $\NH$ and $Z$ which best-fit the observed ion columns in each object, for a total of $3k=132$ free parameters. As discussed in the introduction, this method has been shown repeatedly in the past to fail at reproducing the entire set of absorption features in a given object, since when $\nH$ is optimized on the low-ions, $\ovip$ is typically underpredicted by orders of magnitude. 
The standard alternative is to assume a multi-phase absorber, where $\ovip$ originates from a second phase with a different $\nH$. We therefore use cross-validation also on a two-phase model, which has $5k=220$ free parameters ($\nH^{(1)}$, $\nH^{(2)}$, $\NH^{(1)}$, $\NH^{(2)}$, and $Z$ per object). 
For these constant-density models, we run \cloudy\ with a constant $\log (\nH/\cm^{-3}) = -5.5,\ -5,\ -4.5,\ -4,\ -3.5,\ -3,\ -2.5,\ -2,$ or $-1.5$. All other \cloudy\ parameters are identical to the power-law model described in \S\ref{sec: numerical}. As above, we create a three-dimensional table of predicted ion columns as a function of $Z$, $\nH$, and $\NH$. In the two-phase model, the predicted $\Nion$ is equal to the sum of the predicted $\Nion$ from each phase. 
The third absorber-specific model assumes a power-law density distribution equivalent to that used above, but where each objects has its own distinct profile,
which implies $4k=176$ free parameters ($\dNz, \beta, Z$, and $\rhomax$ per object). 

For each cross-validation iteration, we randomly choose  30\% of the metal ion columns in each object as the validation set.
The free parameters are then optimized
by fitting the training set (the other 70\% of the measurements),
and then the
$\chi^2$ (eqn.~{\ref{eq: likelihood}) of the validation set is calculated. 
A hundred iterations are run for each of the four models, and the $\chi^2$-distributions are shown in Figure~\ref{fig: cross validation}. 

\begin{figure}
\includegraphics{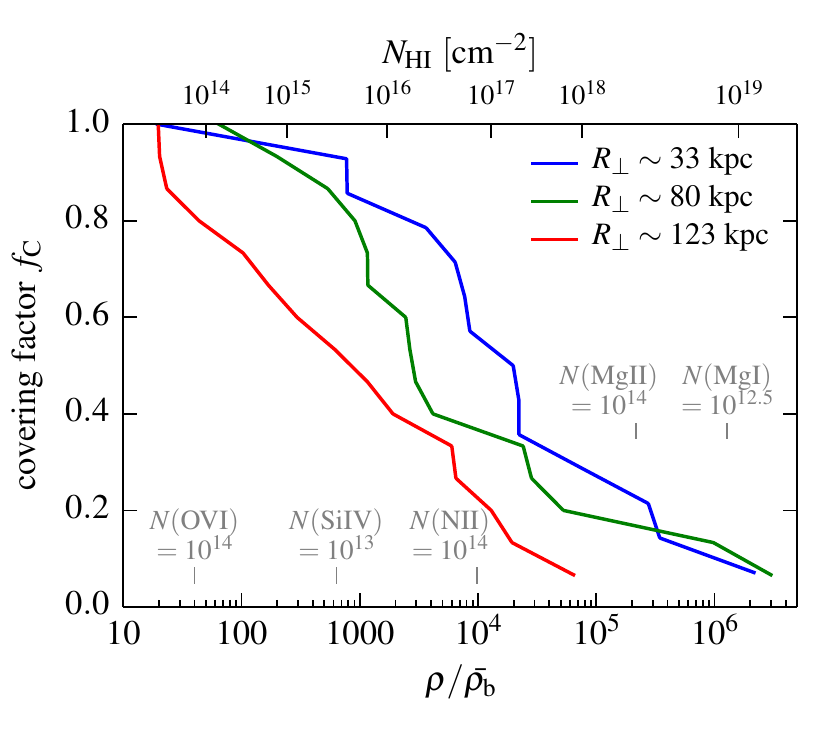}
\caption{
Covering factor versus gas density and impact parameter in the COS-Halos sample. 
In our fit, $\CF(\rho,\Rimp)$ is equal to the fraction of objects with impact parameter $\Rimp$ where the fit $\rhomax$ is larger than $\rho$. 
The plotted $\CF$ combined with the best-fit AMD (top panel of Figure~\ref{fig: AMD}) are used to constrain the number density of clouds at each phase. 
Note that $\CF$ is a weak function of $\rho$, decreasing by $\sim 0.2$ per decade in $\rho$. 
For each $\rho$ we note the associated $\NHI$ on top. 
Small vertical ticks mark $\rho$ where different ions reach the noted columns (see Figure~\ref{fig: fit by ion}), for the typical $Z=0.6\zsun$.
}
\label{fig: CF}
\end{figure}

Figure~\ref{fig: cross validation} shows that the median $\chi^2$ of the universal model used in this study is 
a factor of $\sim$20 lower than the median $\chi^2$ in the absorber-specific constant density model, and
a factor of $\sim$5 lower than in the absorber-specific two-phase model. 
The $\chi^2$-distributions of the universal and absorber-specific power-law models are comparable. 
This result implies that the power-law models
have stronger predictive power than the traditional single and double
density methods used in the literature, and they are therefore favored.

Figure~\ref{fig: cross validation} also shows that the $\chi^2$ distribution of the universal and absorber-specific power-law
models are similar. If the density structure across the COS-Halos sample had been perfectly universal, one would expect the $\chi^2$ distribution of the universal power-law model to be lower than the $\chi^2$ distribution of the absorber-specific power-law model, since the extra flexibility provided by the unnecessary free parameters in the absorber-specific model reduce its predictive power. 
In contrast, if the density structure along sightlines in the COS-Halos sample had a large dispersion, the universal model would produce bad fits to the data and therefore would have a higher $\chi^2$ distribution than the absorber-specific power-law model. 
The similar $\chi^2$ distribution of the two power-law models hence suggests that there is some dispersion among the density structure of the different COS-Halos sightlines, though this dispersion is not very large. This dispersion may be connected to the difference in the CGM of star-forming and quiescent galaxies found by \cite{Tumlinson+11}, or alternatively to the relatively wide range of impact parameters probed ($18\kpc < \Rimp < 154\kpc$).  This dispersion will be explored in future work.

\section{Implied CGM characteristics}\label{sec: CGM characteristics}

In this section we combine the best-fit results of the previous section with the formalism developed in \S\ref{sec: method} in order to calculate aggregate CGM characteristics. 

\subsection{Covering Factors}

In the context of our hierarchical model, $\rhomax$ is the highest density encountered along a given line-of-sight, which thus determines the lowest ionization state along the sightline. 
As such, the covering factor $\CF(\rho,\Rimp)$ is equal to the chance a sightline with impact parameter $\Rimp$ has $\rhomax>\rho$. We therefore divide the sample into three bins in $\Rimp$, namely $18<\Rimp<50\kpc$, $50<\Rimp<100\kpc$, and $100<\Rimp<154\kpc$, each with $14-15$ objects\footnote{The COS-Halos sample was selected so each of these bins has the same number of objects (\citealt{Tumlinson+13}).}. The implied $\CF(\rho,\Rimp)$ are shown in Figure~\ref{fig: CF}. 
The values of $\CF$ drop with increasing $\rho$ (by construction) and with increasing $\Rimp$, from $\CF = 0.8 - 1$ of the low-density $\ovip$ phase to $\CF = 0 - 0.2$ of the high-density $\mgip$ phase.
Note though that the drop in $\CF$ is a weak function of $\rho$ (roughly logarithmic), with $\CF$ dropping by only $\sim0.2$ per decade in $\rho$. 

To associate $\CF(\rho,\Rimp)$ with covering factors of ions, we mark in Figure~\ref{fig: CF} the values of $\rho$ where several ions reach the noted columns, based on the relation between $\rho$ and $\Nion/Z$ plotted in Figure~\ref{fig: fit by ion}, assuming the median of the best-fit metallicities $Z=0.6\zsun$.

\subsection{Cloud Sizes}\label{sec: cloud sizes}

We return to the discrete picture where consecutive values of $\rho$ differ by an order of magnitude ($\rhoi=10\rhoim$). In the appendix we calculate the relation between the fine sampling of $\rho$ in the \cloudy\ calculation used above and the discrete densities $\rhoi$ used here. The implied lowest density phase in the discrete picture of our best-fit model is $\rhoz=50\rhobar$ ($n_{\rm H,0}=1.6\times10^{-5}\cm^{-3}$). 
The best-fit $\beta=0.05$ implies that
the column of this phase is $\dNz(50\rhobar/20\rhobar)^{0.05}\approx\dNz$, where $20\rhobar$ is the minimum density in the \cloudy\ calculation and the best-fit $\dNz$ is given by eqn.~(\ref{eq: AMD fit}).

An initial estimate for the size of clouds with different $\rho$ can be obtained from the relations $\rz=3\dNz/(4n_{\rm H,0})$ and $\alpha=\beta-1$ (eqn.~\ref{eq: AMD vs. density structure values2}), which is accurate in the limit that clouds with the same $\rho$ do not overlap along the line-of-sight. However, as noted above the assumption that same $\rho$ clouds do not overlap is a simplification which over-predicts $\rc$. 
Below we show that when accounting for cloud overlap, the implied clouds sizes are a factor of $\approx1.5$ smaller then the sizes estimated in the no-overlap limit. Hence, $\rz\approx3\dNz/(4n_{\rm H,0})/1.5$, and the best-fit $\dNz$ and $\beta$ (eqn.~\ref{eq: AMD fit}) imply
\begin{equation}\label{eq: density structure values}
 \rc(\rho) = 35\left(\frac{\rho}{50\rhobar}\right)^{-0.95}\kpc ~.
\end{equation}
This relation is shown in the top panel of Figure~\ref{fig: properties by rho}, which also illustrates the errors implied by the bootstrap errors on $\beta$ and $\dNz$ (eqn.~\ref{eq: AMD fit}).
The characteristic size of the densest phase found above ($\rho_4=5\times10^5\rhobar$) is hence $r_{\rm c,4}\approx6\pc$, almost four orders of magnitude smaller then the size of the \ovip-phase. 

\subsection{Filling Factor and Mass}\label{sec: fV}

\begin{figure}
\includegraphics{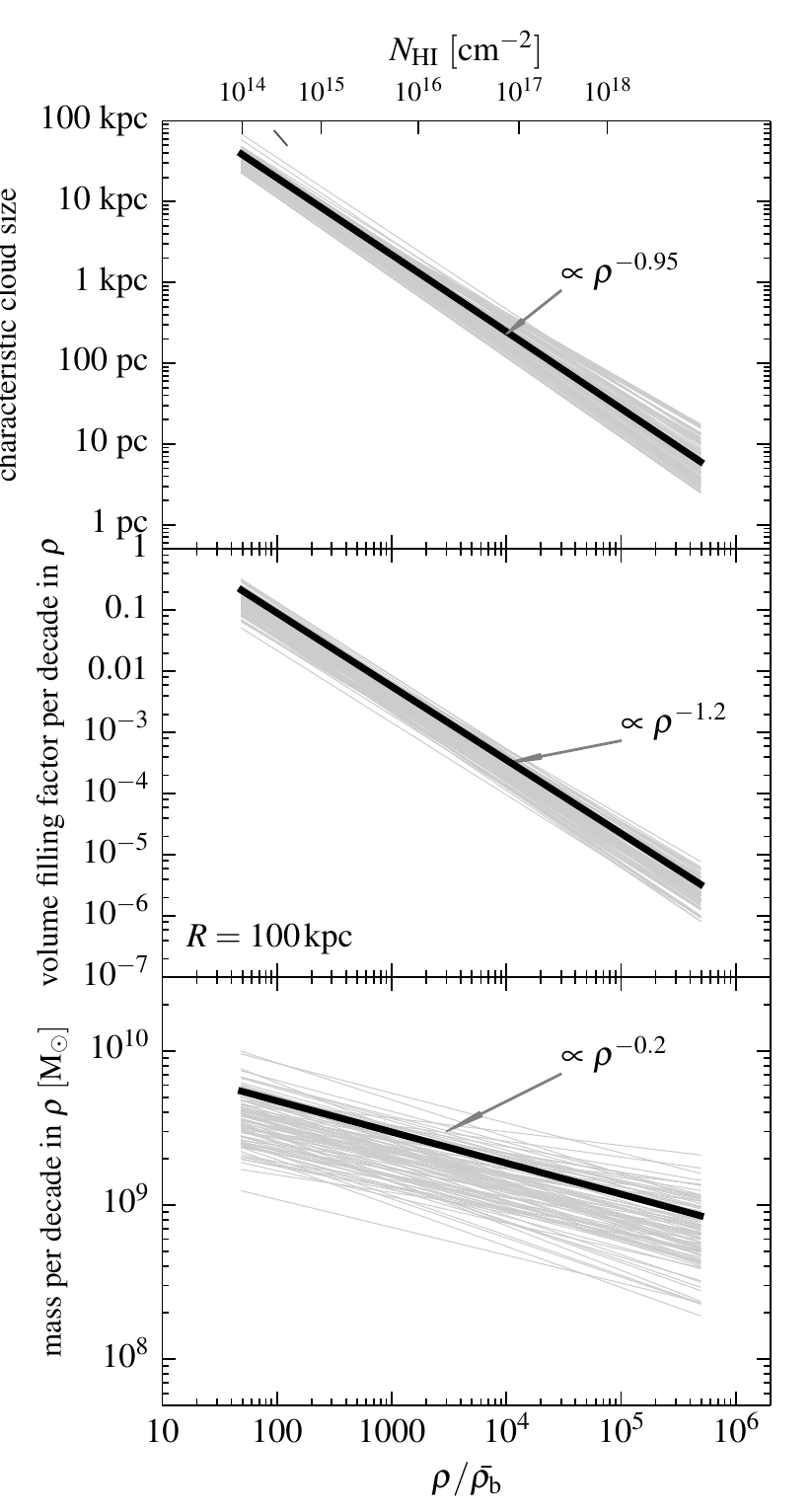}
\caption{
The characteristics of the $T\sim10^4\K$ CGM of COS-Halos galaxies versus gas density, as implied by the best-fit model. 
The distribution of thin gray lines provide an estimate of the error in our procedure using the bootstrap method.
The typical $\NHI$ associated with each $\rho$ is noted on top. 
{\bf (Top)} The relation between characteristic cloud size and gas density (eqn.~\ref{eq: density structure values}). 
{\bf (Middle)} The relation between volume filling factor and gas density at a distance $R=100\kpc$ from the galaxy (eqn.~\ref{eq: fV fit}).
{\bf (Bottom)} The relation between cool CGM mass within $\rvir$ and gas density (eqn.~\ref{eq: dM/drho values}).
}
\label{fig: properties by rho}
\end{figure}

The flat AMD found above implies that the cloud size scales roughly as $\rc \sim \rho^{-1}$ (eqn.~\ref{eq: density structure values}). In contrast, the covering factor depends only logarithmically on $\rho$ (Figure~\ref{fig: CF}). We can use these two properties to gain intuition on the dependence of the filling factor on $\rho$, in the context of the hierarchical model where multiple dense clouds are embedded in a larger low-density cloud.
Since the cross-sectional area of the cloud scales as $\sigc \propto \rc^2$, then for the phases $\rhoim$ and $\rhoi=10\rhoim$ to have the same $\CF$, each $\rhoim$ cloud needs to be populated by $\sim(\rc(\rhoi)/\rc(\rhoim))^{-2}\sim 100$ clouds with density $\rhoi$. 
Since the cloud volume scales as $\rc^3$, these 100 clouds fill a fraction of $100(\rc(\rhoi)/\rci(\rhoim))^3\sim0.1$ of
the parent cloud volume, and have a total gas mass which is roughly equal to the mass of the parent cloud. 
Hence, the weak dependence of $\CF$ on $\rho$ seen in Figure~\ref{fig: CF}, combined with the flat AMD found above, imply roughly equal CGM mass per decade in $\rho$, but a filling factor which scales as $\sim\rho^{-1}$. 

A more accurate calculation of $\fV(\rho,\Rimp)$ can be derived from eqns.~(\ref{eq: NH Rimp avg}) and (\ref{eq: AMD av}), which together give
\begin{eqnarray}\label{eq: fV from CF and PL}
 \CF(\rhoi,\Rimp) \dNz\left(\frac{\rho}{\rhoz}\right)^\beta &=& \langle\AMDfrac(\rhoi,\Rimp)\rangle  \nonumber\\                             &=& \nHi\int \fV(\rhoi,R)\deriv s ~.
\end{eqnarray}
Defining $\mu$ as the cosine of the angle between the plane of the sky and a radial vector to a point along the sightline, 
we get $R(s)=\Rimp/\mu$ and $\deriv s = \Rimp \deriv\mu/(\mu^2\sqrt{1-\mu^2})$. 
Using these geometrical relations in eqn.~(\ref{eq: fV from CF and PL}) we get 
\begin{eqnarray}\label{eq: AMD vs fV developed}
 \CF&(&\rhoi , \Rimp)  \cdot \dNz\left(\frac{\rhoi}{\rhoz}\right)^\beta = \nonumber \\
 & &  \frac{\rhoi}{\mu\mp}\rvir\fVz\left(\frac{\rhoi}{\rhoz}\right)^m\left(\frac{\Rimp}{\rvir}\right)^{l+1} \int_{\frac{\Rimp}{\rvir}}^1 \frac{2\deriv\mu}{\mu^{2+l}\sqrt{1-\mu^2}}  \nonumber \\
\end{eqnarray}
where we used the power-law form for $\fV$ (eqn.~\ref{eq: fV def}), 
and we set the limits of the integral from the edge of a sphere with $R=\rvir$ to the mid-plane, and multiply by two. 

Using least square minimization, we find the $\fVz$, $m$, and $l$ which best-fit the observed $\dNz$, $\beta$, and $\CF(\rhoi,\Rimp)$ according to eqn.~(\ref{eq: AMD vs fV developed}). This process yields
\begin{equation}\label{eq: fV fit}
 \fV(\rhoi,R) = 0.077 \left(\frac{\rhoi}{50\rhobar}\right)^{-1.20} \left(\frac{R}{\rvir}\right)^{-0.97} 
\end{equation}
For deriving equation~(\ref{eq: fV fit}) we assume $\rvir=280\kpc$, the average virial radius of COS-Halos galaxies (\citealt{Werk+14}). Eqn.~(\ref{eq: fV fit}) is plotted in the middle panel of Figure~\ref{fig: properties by rho}. This Figure also illustrates the bootstrapped errors in our calculation, where in each bootstrap iteration we choose with replacements 15 objects in each $\Rimp$-bin to derive $\CF(\rhoi,\Rimp)$, and choose randomly one of the bootstrapped $\dNz$ and $\beta$ shown in the top panel of Figure~\ref{fig: AMD}.

We now use eqn.~(\ref{eq: fV fit}) to derive aggregate CGM characteristics. 
By eqn.~(\ref{eq: dM/drho}), the $T\sim10^4\K$ CGM mass in each density phase is 
\begin{equation}\label{eq: dM/drho values}
 \frac{\deriv\Mcool}{\deriv\log\rho}(<\rvir) = 0.5\times10^{10} \left(\frac{\rhoi}{50\rhobar}\right)^{-0.20}\msun ~.
\end{equation}
Equation~(\ref{eq: dM/drho}) demonstrates that the mass is distributed roughly equally between the different density bins, 
consistent with our simplified estimate above. 
Eqn.~(\ref{eq: dM/drho values}) is plotted in the bottom panel of Figure~\ref{fig: properties by rho}. 

\begin{figure}
\includegraphics{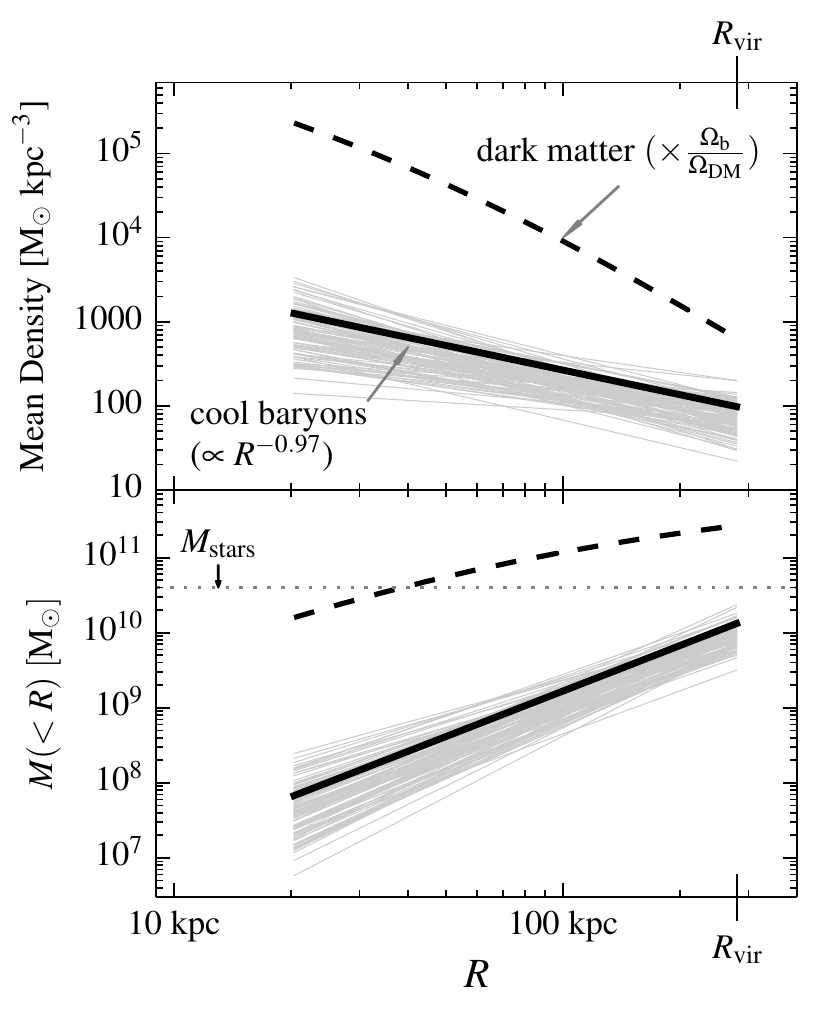}
\caption{
The characteristics of the $T\sim10^4\K$ CGM of COS-Halos galaxies versus distance from the galaxy center, as implied by the best-fit model. 
The distribution of thin gray lines provide an estimate of the error in our procedure using the bootstrap method. 
{\bf (Top)}
The mean density profile (eqn.~\ref{eq: rhomean val}).
The dashed line is the dark matter density profile multiplied by the cosmic baryon mass fraction of $0.17$.
{\bf (Bottom)}
The total mass within $R$ (eqn.~\ref{eq: M(<R) val}).
As in the top panel, the dashed line is the normalized dark matter mass.
The median stellar disk mass is also marked.  
Inside $\rvir$ the cool CGM gas accounts for only $\sim$5\% of the total baryon budget ($=0.17\,M_{\rm halo}$) of an $\sim L^*$ galaxy.
}
\label{fig: properties by R}
\end{figure}

Eqn.~(\ref{eq: fV fit}) also implies that the average density profile of the cool CGM as a function of $R$ is
\begin{equation}\label{eq: rhomean val}
\rhocool(R) = \sum_{50\rhobar}^{5\times10^5\rhobar} \rho \fV(\rhoi,R) 
            = 97\left(\frac{R}{\rvir}\right)^{-0.97} \msun \kpc^{-3} 
\end{equation}
and the total cool gas mass within $R$ is (eqn.~\ref{eq: Mcool})
\begin{equation}\label{eq: M(<R) val}
 \Mcool(<R) = 1.3\times10^{10}\left(\frac{R}{\rvir}\right)^{2.03}\msun ~.\\
\end{equation}
These profiles and their bootstrapped errors are shown in Figure~\ref{fig: properties by R}. 
The marginalized error on the slope of the profile is $\pm0.31$, while the implied total mass within the virial radius is
\begin{equation}\label{eq: Mcgm val}
 \Mcool(<\rvir) = (1.3\pm0.4)\times10^{10}\msun ~.
\end{equation}

\subsection{CGM realization}\label{sec: CGM realization}

\begin{figure*}
\includegraphics{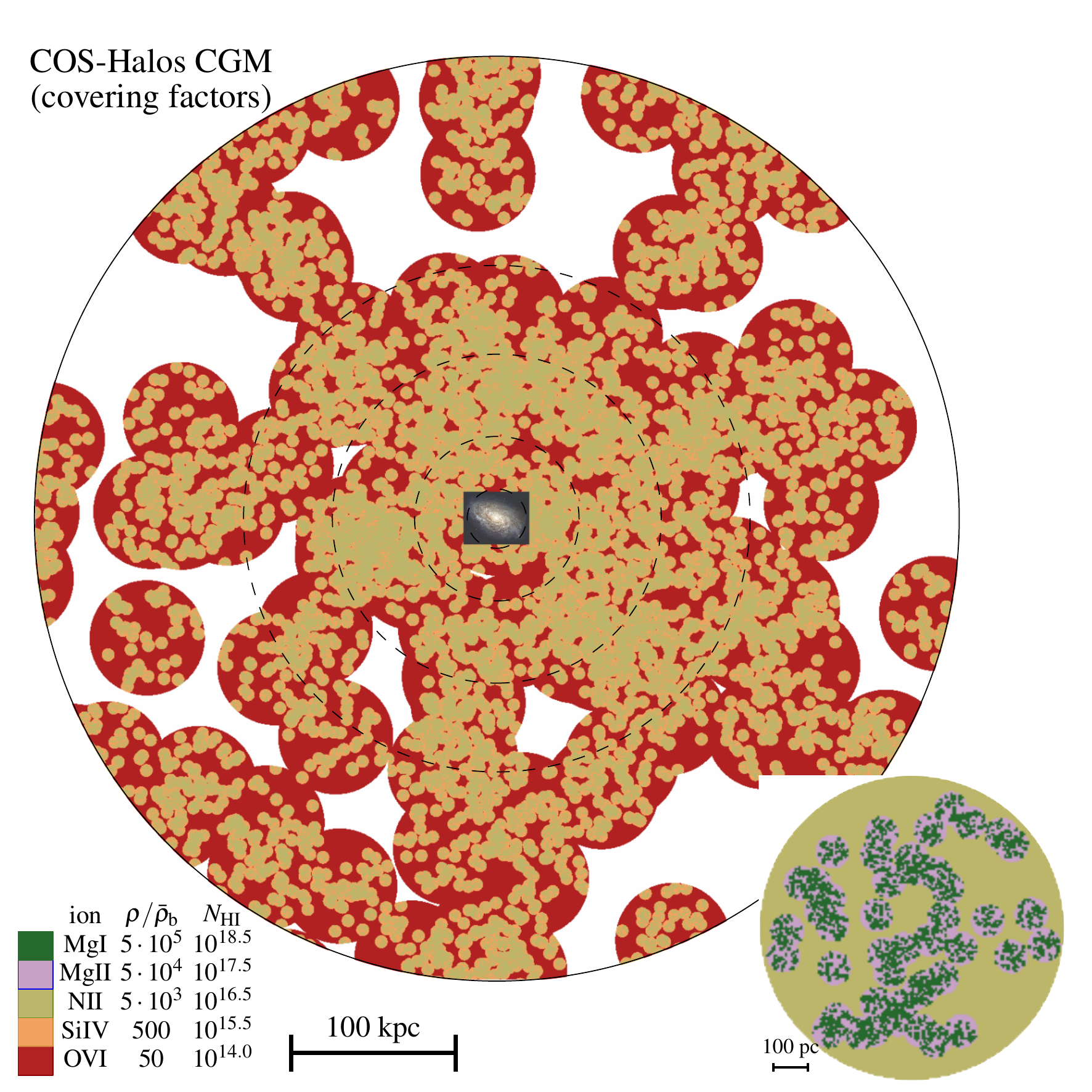}
\caption{
A realization of the cool CGM of COS-Halos galaxies based on the CGM properties shown in Figures~\ref{fig: properties by rho}--\ref{fig: properties by R}.
Characteristic ions and typical $\nhi$ (in $\cm^{-2}$) for each phase are noted in the legend.
The three low-density phases are shown in the main panel, while the inset zooms on a single \niip-phase cloud in order to reveal the two highest-density phases. 
The plot shows the CGM projected on the sky (the line of sight is perpendicular to the paper). Color denotes the maximum density probed by each sightline. The plot demonstrates that the covering factor $\CF$ decreases rather weakly with density. The limits of the three $\Rimp$ bins used in Figure~\ref{fig: CF} are marked by dashed lines. 
}
\label{fig: realization projected}
\end{figure*}

\begin{figure*}
\includegraphics{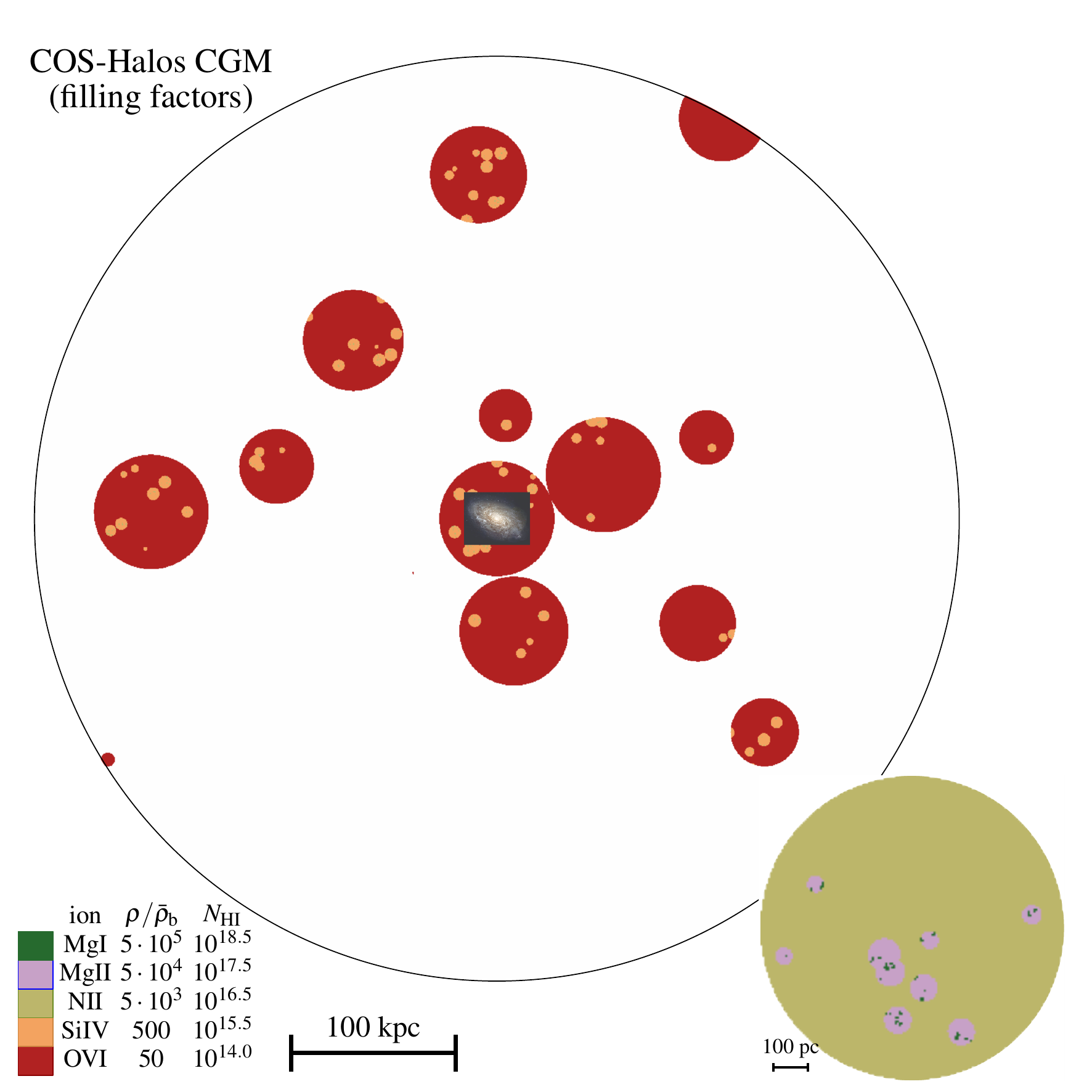}
\caption{
Similar to Figure~\ref{fig: realization projected}, but showing an infinitesimal slice through the mid-plane of the CGM realization. 
Color denotes the gas density in the mid-plane. The plot demonstrates that the volume filling factor $\fV$ decreases strongly with increasing density. 
}
\label{fig: realization slice}
\end{figure*}

To provide a visualization of the hierarchical CGM
structure derived in this study, we create a realization of the CGM based on the physical parameters $\rc(\rhoi)$ and $\fV(\rhoi,\Rimp)$ found above. We use the realization also to justify our claim in \S\ref{sec: cloud sizes} that when accounting for cloud overlap along the line of sight, the implied $\rc(\rhoi)$ are lower by a factor of $1.5$ than when $\rc$ are estimated in the no-overlap limit. 

We randomly populate a sphere with size $\rvir=280\kpc$ with the $i=0$ clouds which have $\rz=35\kpc$. 
The number and distribution of the clouds are set to reproduce the value of $\fV(\rhoz=50\rhobar,R) = 0.08 (R/\rvir)^{-0.97}$ found above (eqn.~\ref{eq: fV fit}). We then randomly populate each $i=0$ cloud with $45$ $i=1$ clouds, in order to reproduce the desired relative filling factor of the two phases of $\fV(\rhoi,R)/\fV(\rhoim,R)=10^{-1.2}=0.06$ (eqn.~\ref{eq: fV fit}), 
given the cloud volume ratio of $(\rc(\rhoi)/\rc(\rhoim))^3=1.4\times10^{-3}$ (eqn.~\ref{eq: density structure values}).
We assume that $i=1$ clouds are uniformly distributed over the volume occupied by the $i=0$ clouds, which creates a separable dependence of $\fV$ on $R$ and on $\rho$, as assumed in eqn.~(\ref{eq: fV def}). 
This recursive populating of clouds is repeated for the $i=2$, $i=3$, and $i=4$ phases.

The CGM realization is shown in Figures~\ref{fig: realization projected} and \ref{fig: realization slice}.
Figure~\ref{fig: realization projected} shows the projection of the CGM on the plane of the sky, where the plotted color denotes the maximum density observed along each line of sight. This figure depicts the covering factors of the different phases plotted in Figure~\ref{fig: CF}.  
Figure~\ref{fig: realization slice} shows an infinitesimal slice of the CGM through the mid-plane, where color denotes the gas density. This Figure depicts the filling factors of the different phases (eqn.~\ref{eq: fV fit}).
Table~\ref{tab: realization} lists for each phase the size and mass of single clouds, and the average number of clouds within a cube with edge size $10\kpc$.

As discussed in \S\ref{sec: hierarchical}, eqn.~(\ref{eq: AMD vs. density structure values2}) overestimates $\rc(\rhoi)$ by a factor equal to the average number of $\rhoi$-clouds along sightlines which intersect at least one $\rhoi$-cloud. 
Table~\ref{tab: realization} lists this factor for each phase, calculated along skewers through the realization that have $18\kpc<\Rimp<154\kpc$, the range of $\Rimp$ probed by the COS-Halos survey. The mean number of clouds are all in the range $1.3-1.7$, within $15\%$ of the factor of $1.5$ used to derive the cloud sizes in eqn.~(\ref{eq: density structure values}), thus justifying the derived sizes. The weak trend in the mean number of clouds with $\rho$ may suggest that low density clouds are somewhat smaller than estimated, while high-density clouds are somewhat larger than estimated. 
The implied change in the index $\alpha$ is however very small, of order $\log(1.7/1.3) / \log(\rho_4/\rhoz) = -0.03$.

The dispersion in the number of clouds listed in the right column of Table~\ref{tab: realization} is
approximately $0.2\dex$ in all phases.
This dispersion is an estimate for the variance in the gas columns of individual sightlines compared to the universal AMD. This result supports our choice above to assume an error of $0.2\dex$ when comparing the universal model predictions with specific ion columns observed in individual sightlines (eqn.~\ref{eq: likelihood}).

The realization also implies that typically $\sim 60\%$ of sightlines which intersect at least one cloud in some phase intersect exactly one cloud of this phase, $\sim 30\%$ intersect two clouds of this phase, and $\sim 10\%$ intersect three clouds. Assuming that different clouds are offset in velocity space, this result can be tested by investigating the number of distinct kinematic components in absorption spectra. We further discuss cloud kinematics in \S\ref{sec: kinematics}.

\section{Discussion}\label{sec: discussion1}

\begin{deluxetable}{cccccc}
\tablecolumns{6}
\tablecaption{Characteristics of each phase in the CGM realization$^{(a)}$\label{tab: realization}}
\tablehead{
\colhead{$\rhoi/\rhobar$} & \colhead{ion} & \colhead{$\rc$}  & \colhead{cloud mass} & \colhead{clouds per}         & \colhead{clouds$^{(b)}$} \\
                          &               &                   & $[\msun]$            & \colhead{$(10\kpc)^{3}$}  & \colhead{per LOS} }
\startdata
            $50$  &   \ovip &   $35\kpc$ & $0.9\times10^8$  & $0.003$  & $1.7$ \\
           $500$  &  \Siivp &  $3.9\kpc$ & $1.3\times10^6$  & $0.12$   & $1.5$ \\
$5 \times 10^{3}$ &   \niip &   $440\pc$ & $1.9\times10^4$  & $5.5$    & $1.4$ \\
$5 \times 10^{4}$ &  \mgiip &    $49\pc$ & $260$            & $240$    & $1.4$ \\
$5 \times 10^{5}$ &   \mgip &   $5.5\pc$ & $3.6$            & $11000$  & $1.3$ 
\enddata
 \tablenotetext{(a)}{The CGM realization is based on the derived cool CGM properties shown in Figures~\ref{fig: properties by rho}--\ref{fig: properties by R}, and is plotted in Figures~\ref{fig: realization projected}--\ref{fig: realization slice}.}
 \tablenotetext{(b)}{Mean number of clouds along skewers through the realization which intersect at least one cloud and have $18<\Rimp<154\kpc$.} 

\end{deluxetable}

In this study, we assume a phenomenological model for the CGM, where the cool photoionized gas is composed of small dense clouds which are hierarchically embedded within larger lower-density clouds, with some characteristic relation between physical scale and gas density.
We develop a method to combine (or `stack') the observations from all 44 COS-Halos objects, and thus yield tight constraints on the density structure. 
This universal phenomenological model produces an acceptable fit to all observed ions, including both the low-ions and the high-ions up to \ovip\ (Figures~\ref{fig: fit by ion}--\ref{fig: fit by obj}), and has both higher predictive power and fewer parameters than the standard models used in the literature (Figure~\ref{fig: cross validation}).
In this section, we discuss some of the uncertainties in our analysis, the implications and predictions of our derived quantities, and how our results can be compared to hydrodynamical simulations.

\subsection{Uncertainty in the ionizing spectrum}\label{sec: UVB}

As shown in Figure~\ref{fig: ionfractions}, the ionization fractions of the different ions are sensitive to $U$, the ratio of the ionizing photon flux $\phii$ to the gas density. Therefore, the range of densities of $50 \lesssim \rho/\rhobar \lesssim 5\times10^5$ derived above depends (linearly) on the assumed $\phii$. We assume above the value of $\phii$ found by HM12, which is based on the luminosity function of quasars at $z=0.2$, and assuming that the contribution from star forming galaxies (and other sources) is negligible. However, recently \cite{Kollmeier+14} compared low-$z$ \Lya\ forest observations with cosmological simulations, and found that the required $\phii$ is higher by a factor of $\sim 5$ than HM12 synthesized by summing the emission from all sources of ionizing photons.
The quantitative value of this discrepancy, however, has been contested by follow-up studies, which found a weaker discrepancy of a factor of $\sim 2$ (\citealt{Shull+15}; Heilker, private communication).
The uncertainty in $\phii$ propagates to our derived $\rho$, e.g.\ if we assume $\phii = 5 \phiHM$, than the implied CGM density range is increased by a factor of five to $250 \lesssim \rho/\rhobar \lesssim 2.5\times 10^6$. 

Also, since the observations constrain the column densities $\AMD \sim \nH \rc(\rho)$,
than any uncertainty in $\rho$ propagates to an uncertainty in $\rc$. Therefore, assuming a \citeauthor{Kollmeier+14}\ UVB implies that the characteristic size of the \ovip-phase is $r_0\approx7\kpc$, rather than the $r_0\approx35\kpc$ derived above (eqn.~\ref{eq: density structure values}).
Reversing the argument, then if we had an independent estimate of the cloud sizes, say by measuring the ion coherence scale (see below), our results would provide a constraint on $\phii$.

On the other hand, the derived CGM mass estimates do not depend on the assumed $\phii$.
This follows since $M \propto \sum \rhoi \fV(\rhoi)$ (eqn.~\ref{eq: Mcool}), and $\rhoi\fV$ is proportional to $\AMD$ (eqn.~\ref{eq: NH Rimp avg}). The CGM mass is essentially an integration of the observed columns over the CGM cross-section (eqn.~\ref{eq: M vs N}). 
Since the hydrogen columns are derived from the observed ionic columns after ionization corrections (eqn.~\ref{eq: nion}), and the ionization corrections are independent of the
absolute density scale which is set by $\phii$, the CGM mass is hence also independent of $\phii$.

Another potential source of ionizing photons is star formation (SF) in the local galaxy (\citealt{Miralda-Escude05, Schaye06}). To estimate the contribution of local SF to the ionizing spectrum we follow \cite{Kannan+14}, who assumed the SED of a $5\Myr$ old stellar population and an escape fraction of $f_{\rm esc}=0.05$ (black curve in fig.~1 there). The ratio of $1\ryd$ photons from the galaxy $\phi_{\rm glx}$ to the HM12 UV background at $z=0.2$ is hence
\begin{equation}\label{eq: SF contribution}
 \frac{\phi_{\rm glx}(1\ryd)}{\phiHM(1\ryd)} = 0.5 \left(\frac{R}{100\kpc}\right)^{-2}\left(\frac{SFR}{\msun\yr^{-1}}\right)\left(\frac{f_{\rm esc}}{0.05}\right)~,
\end{equation}
where the median star formation rate (SFR) in the COS-Halos sample is $1.2\msun\yr^{-1}$ (\citealt{Werk+13}). 
Bearing in mind the high uncertainty in $f_{\rm esc}$, eqn.~(\ref{eq: SF contribution}) suggests a significant contribution to the ionizing flux from local SF, especially in blue galaxies and at low $R$. A significant contribution from the local galaxy would both decrease the implied size of the absorbers as discussed above, and imply a mean cool gas density profile steeper than the $\rhocool \sim R^{-1}$ found in Fig.~\ref{fig: properties by R}. 

The ionizing spectrum may also have a different shape then calculated by HM12, for example a softer spectrum is expected if SF in the galaxy dominates the photon budget. A different spectral shape implies that the ion fractions peak at different $U$ than shown in Figure~\ref{fig: ionfractions}, which will affect the derived density structure. 
We estimate the effect of a different spectral shape by recalculating Fig.~\ref{fig: ionfractions} with a hard ionizing spectrum ($\aion=-0.5$, where $J_\nu \propto \nu^{\aion}$) and a soft ionizing spectrum ($\aion=-2$). For comparison, the spectral slope in HM12 is roughly $-1.5$. 
The peak $U$ of \ovip\ is shifted by a factor of $\sim 3$ to lower values in the hard spectrum, and by a factor of $\sim 3$ to higher values in the soft spectrum. The peak $U$ of lower-ions are shifted by a smaller amount, where the shift generally decreases with decreasing ionization energy, as expected.
A factor of $\sim 3$ change in the peak $U$ implies a factor of $\sim 3$ change in the implied $\rho$ and also in the implied cloud size. However, this change is small compared to the range of $\sim10^4$ in $\rho$ deduced above, suggesting that the uncertainty in the spectral shape does affect our conclusion that the cool CGM spans a large range in gas density.

\subsection{Implications for the velocity field}\label{sec: kinematics}

Figure~11 in \cite{Werk+13} plots the ion absorption profiles seen in 14 COS-Halos objects with detections in both high-ionization and low-ionization lines. The mean velocities of the highly-ionized metal ions are roughly aligned with the mean velocities of low-ionized metal-ions, while the high-ions tend to have broader and smoother profiles than the low-ions. As mentioned in the introduction, these absorption profile characteristics are commonly seen in CGM absorbers.
A common interpretation is that the high-ions originate from a higher temperature collisionally-ionized phase which resides at the interface of the cool low-ion clouds with an external medium (hence the kinematic alignment, e.g.\ \citealt{Simcoe+02,Simcoe+06,Savage+05a,Tripp+08,Tripp+11,Kwak+11,Fox+13,Lehner+14,Crighton+15}).  
However, the behavior of the absorption profiles is also consistent with (and are a primary motivation for) the hierarchical model presented in this study.
Assuming some smooth velocity field in the halo which is dominated by non-thermal motions,
then a large high-ion cloud will span some portion of it, which will set the shape of the high-ion absorption profile. 
A low-ion cloud embedded in the high-ion cloud will span some portion of the velocity field spanned by the high-ion parent cloud. 
Hence, we expect the kinematics of the low-ion clouds to be some `fraction' of the kinematics of the high-ion parent cloud. 
In other words, the spatial hierarchy assumed in this study implies also a kinematic hierarchy, which means that the low-ion absorption profiles should be roughly aligned and narrower than the high-ion profiles, as observed.

In cases where the line of sight traverses several low-ion clouds (up to three are expected in a typical sightline, see \S\ref{sec: CGM realization}), we expect the low-ion absorption profiles to occupy different locations (in velocity space) within the \ovip\ absorption profile, as is indeed seen in J1016+4706 (fig.~11 in \citealt{Werk+13}). 

A possible challenge for the hierarchical model is the tight kinematic correspondence between low- and intermediate-ions (e.g. $\Siiiip$ vs.\ $\Siiip$, $\ciiip$ vs.\ $\ciip$) noted by \cite{Werk+13}. In the hierarchical model these ions are not entirely co-spatial, and hence the kinematic alignment should not be perfect. We defer a thorough analysis of the velocity profiles of the ions in the context of the hierarchical model to future work.

The expected \hi\ profile in our picture is somewhat more complex, since the kinematically-broad \ovip\ clouds are associated with a characteristic $\nhi$ of $\sim10^{15}\cm^{-2}$, while the kinematically-narrow low-ion clouds are associated with larger $\nhi$. Therefore, the total \hi\ profile should appear as the sum of these different components. 
A possible test of the model is using lines of sight with $\nhi \sim 10^{15}\cm^{-2}$, which should intersect an \ovip\ cloud but no low-ion clouds. In this case, there should be no confusion with \hi\ absorption from the low-ion phases, and the \hi\ absorption should originate from the \ovip-absorbing gas. Since in our photoionized picture the temperature is low ($\sim30\,000\K$) even in the \ovip-phase, the contribution of thermal broadening $b_{\rm T}$ to the total broadening $b$ is small, $b_{\rm T}(\hi)\sim20\kms$ and $b_{\rm T}(\ovip)\sim5\kms$ compared to the median $b(\ovip)=43\kms$ in the COS-Halos sample. Hence, along $\nhi \sim 10^{15}\cm^{-2}$ sightlines we expect the \hi\ velocity profile to be similar to the velocity profile of \ovip. 

\subsection{The ionization mechanism of \ovip}

As discussed in the introduction, the question whether \ovip\ absorption originates in photoionized or collisionally-ionized gas has important implications for the physical conditions in galaxy halos. This question has been addressed by numerous studies using both observational and theoretical arguments (\citealt{Heckman+02,Fox+07,ThomChen08,Tripp+08,Howk+09,WakkerSavage09,OppenheimerDave09,Savage+10,Savage+11a,Savage+11b,Savage+12,Savage+14,Narayanan+10a,Narayanan+10b,Narayanan+11,Narayanan+12,Prochaska+11,Fox11,Bordoloi+16,Gutcke+16,Oppenheimer+16,Fielding+16}).
In this work we assume \ovip\ originates in the lowest-density phase of the photoionized hierarchical structure (Figs.~\ref{fig: realization projected}-\ref{fig: realization slice}). To allow for the possibility that \ovip\ originates instead in collisionally-ionized gas, we refit the free parameters in our model using all COS-Halos data excluding \ovip\ and \nvp. We find $\dNz=10^{18.54}\cm^{-2}$, $\beta=0.08$, and $Z/\zsun=0.5 \pm 0.4 \dex$, 
consistent with the results found above (eqn.~\ref{eq: AMD fit}) when the measured \ovip\ and \nvp\ columns are included in the fit . Thus, if \ovip\ originates in collisionally-ionized gas, then the cool CGM picture we derive for the mid- and low-ions remains intact, and one must only exclude the lowest-density phase (i.e., the red circles in Figs.~\ref{fig: realization projected}-\ref{fig: realization slice}). 

We note that the success of our fit provides a challenge for scenarios where \ovip\ is collisionally-ionized. 
Since we derive the same best-fit parameters when we exclude the \ovip\ measurements, our model essentially predicts the typical \ovip\ column based on the typical columns of the mid- and low-ions. That is, in our model where density is a smoothly varying parameter, the higher density phases connect smoothly via our two-parameter density profile to the low-density phase probed by OVI. 
If OVI is actually collisionally ionized, then the ability of our model to explain OVI so well is a peculiar coincidence.

Additionally, the photoionized scenario predicts the minimun $\NHI$ associated with $\ovip$ detections. This follows since in our model \ovip\ originates from gas with $U\approx0.03$ in which $f_\ovip$ is maximized (Fig.~\ref{fig: ionfractions}). In these conditions, $N_\ovip/N_\hi \approx 1.1 (Z/Z_\odot)$, which for the deduced $Z\approx0.5\zsun$ and the typical observed $\novi=10^{14.5}\cm^{-2}$ implies $\NHI\approx10^{14.3}\cm^{-2}$. 
The \ovip-panel in Fig.~\ref{fig: fit by ion} shows that the COS-Halos observations are consistent with this prediction -- all objects with \ovip\ detections have $\NHI>10^{14.3}\cm^{-2}$ and all six objects with $\NHI\lesssim10^{14.3}\cm^{-2}$ are not detected in \ovip.  Again, any scenario where \ovip\ is collisionally ionized needs to reproduce this relation between $\NHI$ and $\novi$. 

Another argument invoked to support the collisionally-ionized scenario for \ovip\ is that the line width $b_\ovip$ increases with $N_\ovip$, as expected in a cooling flow where \ovip\ is collisionally-ionized (\citealt{Heckman+02, Sembach+03,Fox11,Bordoloi+16}). We note that the $b_\ovip$ vs.\ $N_\ovip$ trend is also qualitatively expected in the cool CGM picture presented in this work. 
Since the \ovip\ pathlengths of tens of kpc deduced above (Table~\ref{tab: realization}) are a significant fraction of the halo size ($\sim300\kpc$), larger $\novi$ suggest a larger pathlength which samples a larger fraction of the halo gravitational velocity field. Hence, an \ovip\ feature with large $\novi$ is expected to have also a large $b_\ovip$, as observed. 
We defer a quantitative comparison between the observed $b_\ovip$ vs.\ $N_\ovip$ relation and our model to future work.

Based on the detection of broad \Lya\ absorbers (BLAs), \citeauthor{Savage+14} (2014, hereafter S14) found that $31\%$ of \ovip\ absorbers in a blind low-$z$ have $\log T/k\sim5-6$, suggesting \ovip\ originates in warm collisionally ionized gas, rather than in cool photoionized gas as assumed here. As noted by S14, this conclusion depends on the assumption that the BLA and \ovip\ absorption arise in the same gas, while in principle BLAs could also arise in cool gas with large non-thermal broadening which is unassociated with \ovip\ (see also \citealt{Tepper-Garcia+12}). Mass considerations suggest the conditions deduced for the $\log T/k>5$ objects in S14 are unlikely to be applicable to COS-Halos objects. The median $\NH/\novi$ in these objects is $10^{5.7}$ (table~4 in S14). Given the $\novi\approx10^{14.5}\cm^{-2}$ found nearly ubiquitously out to $\rvir$ in blue, COS-Halos-like galaxies (\citealt{Johnson+15}), this $\NH/\novi$ implies an \ovip-gas mass of $5\times10^{11}\msun$ (eqn.~\ref{eq: M vs N}). This mass is larger than the entire baryonic budget of $(\Omega_{\rm b}/\Omega_{\rm DM})M_{\rm halo}=1.5\times10^{11}\msun$ of a blue COS-Halos galaxy, which is unlikely.

\subsection{Prediction for the ion coherence scale}\label{sec: coherence}

A prominent feature of the results of our modeling is the large dynamical range of sizes spanned by CGM clouds, from the $\approx 35\kpc$ 
\ovip-clouds to the $\approx 6\pc$ size of the densest
phase which produces \mgip\ (top panel of Figure~\ref{fig: properties by rho}, Figures~\ref{fig: realization projected}--\ref{fig: realization slice}). Multiple previous studies have already noted that low-ion CGM clouds have sizes of 10s--100s of pc (\citealt{Rauch+99, Prochaska99, Petitjean+00, Rigby+02, Simcoe+06, Schaye+07, ProchaskaHennawi09, RogersonHall12, Stocke+13, Werk+14, Crighton+15, Lau+15}),
thus our results are consistent with these previous results, and extend them by deducing also the larger sizes of the mid- and high-ion clouds.

A testable quantitative prediction of the relation between density and size (eqn.~{\ref{eq: density structure values}) is the coherence scale of each ion. 
Given two lines of sight with some transverse separation, we expect ions which originate from clouds larger than the transverse separation to show a similar absorption profile in both lines of sight, while ions with sizes smaller than the transverse separation to differ in their absorption profile. 
Furthermore, since in our picture small dense clouds are grouped within larger low density clouds, some degree of coherence is expected even at transverse separations larger than the characteristic size associated with the ion, though this coherence should not be perfect, and should decrease with increasing separation. 

The coherence scale can be measured with observations of absorption systems along multiple sightlines with small transverse separations. 
Such sightline pairs are available in samples of gravitationally lensed quasars, in which the transverse separations range from $\sim$10$\kpc$ down to $\sim$10$\pc$ near the redshift of the source where the light paths converge (\citealt{Rauch+99, Rauch+01, Rauch+02, Churchill+03, Ellison+04, Lopez+07, Chen+14}). These studies have deduced a CGM picture consistent with that found here, with $\gg$kpc \ovip\ clouds (\citealt{Lopez+07}), $\sim$kpc \civp\ clouds (\citealt{Rauch+01,Ellison+04,Lopez+07}), and $\lesssim$100$\pc$ low-ionization clouds (\citealt{Rauch+99, Rauch+02,Churchill+03}).
Larger transverse separations can be probed with samples of binary quasars and projected pairs (\citealt{Hennawi+06b, Hennawi+10, Rorai+13, Rubin+15}),
and were suggested as a similar probe of the coherence length of Lyman-Limit Systems (\citealt{Fumagalli+14}). 
A more detailed comparison between our model and coherence scale measurements is deferred to future work.

\subsection{A possible $\rho\approx 5\rhobar$ \neviiip\ phase}\label{sec: neviii}

As mentioned in \S\ref{sec: numerical}, assuming an initial density in the \cloudy\ calculation which is lower than the assumed $20\rhobar$ does not affect the best-fit, since the additional layer of gas is so highly ionized that the contribution to all COS-Halos ions is negligible.
Therefore, based on the COS-Halos observations alone we cannot exclude or detect a photoionized gas phase with $\rho<20\rhobar$. Such a phase can in principle exist at $R\sim\rvir$ where the volume filling factor of all phases discussed above are significantly less than unity (eqn.~\ref{eq: fV fit}).

This phase could however be detected via its \neviiip\ or $\mgxp$ absorption, which exist in appreciable quantities in gas with $\rho\approx5\rhobar$ (see Figure~\ref{fig: ionfractions}). 
By extending the best-fit AMD (eqn.~\ref{eq: AMD fit}) to $\rho=5\rhobar$ we can derive the expected \neviiip\ column from this phase:
\begin{equation}\label{eq: neviii}
 N_{\neviiip} \approx 4\times 10^{13}\left(\frac{\AMDfrac (\rho=5\rhobar)}{10^{18.5}\cm^{-2}}\right) 
\left(\frac{Z}{0.6\zsun}\right) \cm^{-2}~.
\end{equation}
\neviiip\ is detectable with COS at $0.45 < z \lesssim 1$, where it is redshifted to observable wavelengths but $z$ is not too high such that most of the flux is absorbed by Lyman-limit systems. This redshift range has been observed by the CASBaH sample (PI: Tripp), of which first results are published in \citeauthor{Meiring+13} (2013, hereafter M13). M13 find three absorption systems with $0.68 < z < 0.73$, which they attribute to the CGM of $\sim L^*$ galaxies, similar to the COS-Halos sample. The \neviiip\ columns that they find are $9\times 10^{13}$, $7\times 10^{13}$, and $6\times 10^{13}\cm^{-2}$. 
A few additional \neviiip\ systems with similar columns were previously detected by \cite{Savage+05b}, \cite{Narayanan+09,Narayanan+11} and \cite{Hussain+15}.
The observed columns are all within a factor of $2-2.5$ of the value predicted by eqn.~(\ref{eq: neviii}). Therefore, the photoionized gas model deduced in this work predicts the observed \neviiip\ column to within a factor of a few. This success may imply that \neviiip\  originates from low-density photoionized gas.

What are the properties of the \neviiip\ phase, if it is photoionized as suggested by the extrapolation of our model to lower density? 
Since the ionization fraction of a given ion depends on $U$, 
the relation between ionization fraction and $\rho/\rhobar$ depicted in Figure~\ref{fig: ionfractions} depends on the evolution of the quantity $\phii / \rhobar$ with $z$. 
\cite{Shull+12} showed that $\phiHM(z)\propto (1+z)^{4.4}$ at $0<z<0.7$, which combined with $\rhobar \propto (1+z)^{3}$ implies $\phii / \rhobar$ is 60\% larger at $z=0.7$ than at $z=0.2$. 
Therefore, for a HM12 background \neviiip\ at $z=0.7$ traces gas with $\rho\approx8\rhobar$. 
The implied characteristic size of the \neviiip\ phase at $z=0.7$ is hence (eqn.~\ref{eq: AMD vs. density structure values1})
\begin{equation}\label{eq: r_NeVIII}
 \rc(8\rhobar) = \frac{3\AMDfrac(\rho=8\rhobar)}{4\cdot8\,\nHbar(z=0.7)} = 100\kpc ~,
\end{equation}
where we used $[\AMD](\rho=8\rhobar)=10^{18.5}\cm^{-2}$ and $\nHbar(z=0.7)=10^{-6}\cm^{-3}$. 
Given that our \neviiip\ column predictions underestimate the observed columns by a factor of $\sim2$, $\rc$ may also be underpredicted by a factor of two, implying that $\rc(8\rhobar)\sim 200\kpc$, i.e.\ the \neviiip\ phase fills most of the halo of $L_*$ galaxies. 
Note that the size derived in eqn.~(\ref{eq: r_NeVIII}) is significantly lower than 11\Mpc, the pathlength estimated by \cite{Savage+05b} by assuming \neviiip\ is photoionized. The difference is mainly because \citeauthor{Savage+05b}\ required the observed $\nhi=10^{15}\cm^{-2}$ to originate in the same gas as \neviiip. In our hierarchical model the \neviiip-gas has $\NH\approx10^{18.5}\cm^{-2}$ and $f_\hi=10^{-5.2}$ (Fig.~\ref{fig: ionfractions}), which implies $\nhi=10^{13.3}\cm^{-2}$. The larger observed $\nhi$ originates in denser gas. 

To conclude this section, extrapolating our model to lower density
suggests that the halos of $\sim L^*$ galaxies at $z\approx 0.7$ are
filled with $T\approx 60\,000\K$ gas (the temperature of photoionized
gas with $\log U=-0.5$), which is metal-enriched and has a density
eight times the cosmic mean.  The expected \neviiip\ absorption from
this phase is consistent with the observations of M13. 
Hence if NeVIII is indeed photoionized according to our model, then
NeVIII absorption systems trace the largest metal enriched regions
around galaxies at gas densities comparable to those in the IGM.
We note though that if the UV background intensity is
stronger than calculated by HM12, as discussed in \S\ref{sec: UVB},
than the density traced by \neviiip\ is correspondingly larger, and
$\rc(\neviiip)$ is correspondingly smaller.

\subsection{CGM mass}

In equation~(\ref{eq: Mcgm val}), we derive a total cool gas mass within the virial radius of $\Mcool=(1.3\pm0.4)\times10^{10}\msun$. For comparison, \citeauthor{Werk+14} (2014, hereafter W14) measured the photoionized gas mass in COS-Halos galaxies and found $\Mcool \sim 6.5\times10^{10}\msun$, with a lower limit of $2.1\times10^{10}\msun$.
The difference between the mass estimate here and in W14 is further enhanced by the fact that here we include the \ovip-phase, which accounts for 40\% of the total mass ($M_{{\rm cool},0}=5\times10^{9}\msun$, eqn.~\ref{eq: dM/drho values}), while the W14 estimate does not include the \ovip-phase.

The higher gas mass is due to the higher ionization-corrected $\NH$ deduced by W14. The difference in the 
columns deduced by W14 likely reflect the difference between modeling absorption line data with a constant density model as done by W14, compared to modeling using a multi-density absorber as done here. 
In a multi-density model each ion $\ion$ originates mainly from gas in which $\fion$ peaks, and hence the required gas column $\NH$ required to produce the observed $\Nion\propto\NH\fion$ is minimal. 
In contrast, in a single-density model $\fion$ will inevitably be below-maximal for some of the ions, and hence the fit will tend to deduce larger $\NH$.
The higher predictive power of the power-law models compared to the constant-density models (Figure~\ref{fig: cross validation}) suggests that CGM absorbers are indeed multi-density. 

We also note that the $30\%$ statistical uncertainty in the estimate of $\Mcool$ found here is substantially lower than the factor of $\sim$3 uncertainty deduced by W14. The major source of uncertainty in the W14 analysis is the unknown \hi\ column of objects where the Lyman features are saturated, which compose half of the COS-Halos sample. 
In the constant-density models used by W14, an unknown $\NHI$ implies that $Z$ and $\NH$ are degenerate, since the expected metal columns are roughly proportional to both properties, while there is no independent estimate of the hydrogen column that can break the degeneracy. Hence, $\NH$ (and $\Mcool$) are not tightly constrained in these objects.
In the universal power-law model however, for a given AMD the value of $\NHI$ depends only on $\rhomax$ (eqn.~\ref{eq: nhi by rhomax values}), which in turn sets the lowest ionization level expected in the absorber (see Figure~\ref{fig: fit by ion}). 
Since the universal AMD is constrained by all objects, 
one can calculate the expected $\NHI$ in a given object directly from the lowest-ionization metal ions seen in the absorber, i.e.\ from the fit $\rhomax$. Hence, in these objects
there is no degeneracy between $Z$ and $\NH$, which propagates to a substantially lower uncertainty in the mass estimate. 
Thus, the significantly lower uncertainty on $\Mcool$ derived in this work compared to previous estimates demonstrates the advantage of using our modeling technique for CGM absorbers.

The estimate of $\Mcool$ found in this work suggests that within $\rvir$, the cool gas mass is only $30\%$ of the typical stellar disk mass of $4\times10^{10}\msun$ in COS-Halos galaxies (\citealt{Werk+12}). Adding an unobserved $\rho=5\rhobar$ phase, as suggested by observations of \neviiip\ (\S\ref{sec: neviii}), will increase $\Mcool$ by $0.8\times10^{10}\msun$ (eqn.~\ref{eq: dM/drho values}) to half the stellar disk mass. Thus, our results suggest that including the cool CGM baryons does not significantly change the total baryon content of the galaxy.

Figure~\ref{fig: properties by R} compares the cool gas density and mass profiles (eqns.~\ref{eq: rhomean val}--\ref{eq: M(<R) val}) with the profiles implied by multiplying the
dark matter profile (\citealt{Navarro+97}) by the cosmic baryon mass fraction of $\Omega_{\rm b} / \Omega_{\rm DM}=0.17$. The dark matter profile is calculated assuming $M_{\rm halo}=1.6\times10^{12}\msun$, the median halo mass in the COS-Halos sample (\citealt{Werk+14}), and a concentration parameter of $8$ (\citealt{DuttonMaccio14}). 
The top panel demonstrates that the cool baryon density profile we find is significantly flatter than the dark matter profile at CGM scales.
The bottom panel shows that within $\rvir$, the cool CGM baryons account for only $\sim$5\% of the total baryon budget of $0.17\,\Mhalo=2.7\times10^{11}\msun$. 

Figure~\ref{fig: properties by R} demonstrates that $\Mcool(<R)$ increases quadratically with $R$, which may suggest the existence of a significant baryon reservoir at $R>\rvir$. This possibility can be tested by applying our methodology to samples of galaxy-quasar pairs with larger impact parameters than in COS-Halos galaxies, such as \cite{Johnson+15}.

\subsection{What physical process gives rise to the cool CGM density structure?}

Above we find a range of $10^4$ in $\rho$, from the $\rhoz=50\rhobar$ \ovip\ phase up to the $\rho_4=5\times10^5\rhobar$ low-ion phase. 
Since in our model all the gas is photoionized, the gas temperature $T$  decreases only mildly with increasing density, from $30\,000\K$ in the \ovip\ phase down to $6000\K$ in the densest phase. The thermal pressure of the dense clouds is hence $2000$ times larger than the thermal pressure of the \ovip\ clouds. 
In this section we compare the deduced density structure with several simple hydrostatic solutions, and show that they are all unsatisfactory. It is therefore likely that the density structure originates from a hydrodynamic process, the nature of which is currently an open question.

\subsubsection{Confinement by hot gas}\label{sec: hot phase}

While the different phases are clearly not in thermal pressure equilibrium, the outer \ovip\ phase may in principle be in pressure equilibrium with an external hot gas phase. 
In the picture of \cite{MoMiraldaEscude96} and \cite{MallerBullock04}, the hot gas is at the virial temperature ($\Thot\sim10^6\K$) and has an over-density that follows the dark matter over-density ($\approx80\rhobar$ at $\rvir$ for the profile shown in Figure~\ref{fig: properties by R}). Such a scenario has recently been analyzed analytically by \cite{Faerman+16}. 
For the \ovip\ phase to be in pressure equilibrium with this hot phase, the \ovip\ density needs to have a  density of $\sim10^6\K/(3\times10^4\K) \cdot 80\rhobar=2700\rhobar$. 
This estimate is a factor of $50$ higher than the density $50\rhobar$ of the \ovip\ phase deduced here, significantly larger than the factor of a few uncertainty in $\rho(\ovip)$ due to the uncertainty in the ionizing background (\S\ref{sec: UVB}). This discrepancy is even larger at smaller $R$ where the dark matter over-density is higher (Figure~\ref{fig: properties by R}). Hence if the \ovip\ phase is photoionized as our model suggests, 
it is unlikely that that the \ovip-phase is in pressure equilibrium with such a hot gas phase.

\subsubsection{Self-gravity}

Gravity balances pressure at the Jeans scale, hence if the CGM clouds are self-gravitating we expect a cloud size of (\citealt{Schaye01})
\begin{equation}\label{eq: self-gravity}
 r_\text{self-gravity}(\rho) \sim 300 \left(\frac{\rho}{50\rhobar}\right)^{-1/2} \fg^{1/2} \left(\frac{T}{10^4\K}\right)^{1/2}\kpc
\end{equation}
where $\fg$ is the ratio of gas mass to the total gravitating mass (gas, dark matter and stars). 
Note that eqn.~(\ref{eq: self-gravity}) is derived assuming a constant density cloud, while in our picture each parent cloud is populated by higher-density clouds. However, for the mass distribution deduced above where the mass of embedded clouds is roughly equal to the mass of the parent cloud (\S\ref{sec: fV}), the implied correction to $r_\text{self-gravity}$ is small. 
For $\fg=1$, $r_\text{self-gravity}$ implied by eqn.~(\ref{eq: self-gravity}) are larger than the values of $\rc$
derived in eqn.~(\ref{eq: density structure values}), by an order of magnitude for the \ovip-phase, and by three orders of magnitude for the densest phase ($\rho_4=5\times10^{5}\rhobar$), ruling out self-gravity with $\fg=1$.
Alternatively, setting $r_\text{self-gravity}=\rc(\rhoi)$ and solving eqn.~(\ref{eq: self-gravity}) for $\fg$ gives
\begin{equation}\label{eq: fg}
 \fg(\rho) = 0.014 \left(\frac{\rho}{50\rhobar}\right)^{-0.9} \left(\frac{T}{10^4\K}\right)^{-1}
\end{equation}
For the gas mass deduced above for each phase (lower panel of Figure~\ref{fig: properties by rho}, eqn.~\ref{eq: dM/drho values}), the values of $\fg$ in eqn.~(\ref{eq: fg}) imply a total gravitating mass of $1.1\times10^{12}\msun$ for the \ovip\ phase, and significantly larger gravitating masses for the denser phases. Given the expected total halo mass of $10^{12}\msun$, self-gravity is hence ruled out for all phases denser then the \ovip\ phase, and is possible for the \ovip\ phase only if all the halo mass is in the space occupied by the \ovip\ clouds, which is unlikely. 
A similar conclusion was reached by \cite{Simcoe+06} and \cite{Schaye+07} based on the size they deduced for (single phase) high-$z$ absorbers. 

Self-gravity is also disfavored for the dense phases since the required number of mini-halos exceeds the number of mini-halos predicted in cosmological simulations (\citealt{Tumlinson+13}).

\subsubsection{Radiation Pressure Confinement}

Another potential quasi-static solution which produces a density gradient within the absorber is radiation pressure confinement (RPC), where the gas pressure is in equilibrium with the pressure of the absorbed radiation:
\begin{equation}\label{eq: RPC}
 P_{\rm gas;\,RPC} = \frac{1}{c}\int F_\nu (1-e^{-\tau_\nu}) \deriv\nu ~,
\end{equation}
where $F_\nu$ is the flux density and $\tau_\nu$ is the optical depth. 
RPC conditions have been shown to apply in at least some star forming regions (\citealt{Draine11, YehMatzner12, Yeh+13, Verdolini+13}), in AGN emission line regions (\citealt{Dopita+02, Baskin+14a, Stern+14a, Stern+16}), in AGN absorption line regions (\citealt{Stern+14b, Baskin+14b}), and possibly in the CGM of quasar hosts, in the part exposed to the quasar radiation (\citealt{Arrigoni-Battaia+16}). 
Assuming that dust grains are embedded in the CGM gas, as suggested by the results of \cite{Menard+10}, then the dominant contribution to the integral in eqn.~(\ref{eq: RPC}) comes from $\sim1\mic$ photons emitted by the galaxy which are absorbed by the grains.
Therefore,
\begin{equation}
 2n_{\rm H;\,RPC} k T \approx \sigma_{1\mic}\NH \cdot \frac{L_{\rm galaxy}}{4\pi R^2 c}  ~,
\end{equation}
where $\sigma_{1\mic}$ is the dust cross section per H-atom at $1\mic$. 
For a Galactic grain mixture and dust-to-gas ratio we get
\begin{eqnarray}\label{eq: RPC res}
 n_{\rm H;\,RPC} =   1.2\times & & 10^{-6}\frac{L_{\rm galaxy}}{3\times10^{10}L_\odot} \left(\frac{R}{100\kpc}\right)^{-2}  \nonumber\\
& & \cdot\frac{\NH}{10^{19}\cm^{-2}}\frac{\sigma_{d;\,1\mic}}{10^{-22}\cm^2} \left(\frac{T}{10^4\K}\right)^{-1} \cm^{-3} \nonumber\\
\end{eqnarray}
which falls short by an order of magnitude even for the lowest densities found above ($n_{\rm H,0}=1.7\times 10^{-5}\cm^{-3}$). Hence, radiation pressure is too weak to produce the density structure deduced above.

\subsection{Comparison with Hydrodynamical Simulations}

Our derived CGM density distribution can be compared to theoretical predictions of hydrodynamical simulations of $\approx 10^{12}\msun$ halos at $z\sim 0$. The simplest approach would be to compare the deduced filling factors ($\fV(\rhoi,R)$, eqn.~\ref{eq: fV fit}) and mass distribution ($\deriv \Mcool/\deriv\log\rho$, eqn.~\ref{eq: dM/drho values}) with the same parameters in the simulation. 

One can also compare the predictions of the simulations with our deduced AMD ($\deriv\NH/\deriv\log\rho$, eqn.~\ref{eq: AMD fit}) and covering factor distribution (Figure~\ref{fig: CF}), which are the properties we deduce that include the minimal set of assumptions. 
To perform such a comparison, one should draw skewers through halos at impact parameters $30<\Rimp<150\kpc$, isolate the photoionized gas ($T < 3\times10^4\K$), 
and record $\rho$ along each pixel of the skewer. 
Then, for each decade in $\rho$ and for each $\Rimp$, calculate the average AMD   $\langle\deriv\NH/\deriv\log\rho\rangle$ and covering factor $\CF$ in the different skewers. 
According to eqn.~(\ref{eq: AMD av}), the ratio of these two values is equal to $\dNz(\rhoi/\rho)^\beta$, and hence the calculated ratio can be compared to the $\dNz$ and $\beta$ found here (top panel of Figure~\ref{fig: AMD} and eqn.~\ref{eq: AMD fit}).

We note that it is currently challenging for hydro simulations to resolve the dense CGM phases, as discussed in the context of single-phase absorbers in \citeauthor{Crighton+15}\ (2015, \S5.3 there). Zoomed in SPH simulations such as ERIS2 (\citealt{Shen+13}) and FIRE (\citealt{Hopkins+14}) have a particle mass of $2\times10^4\msun$ and $5\times10^3\msun$, respectively. According to Table~\ref{tab: realization}, this particle mass is larger than the masses of single clouds in the two densest phases. If we further apply the requirement of a few thousand particles per cloud in order to resolve hydrodynamic instabilities (\citealt{Agertz+07, Crighton+15}), than the third-densest phase is also not resolvable with current simulations. These three phases make up a third of the total cool gas mass (eqn.~\ref{eq: dM/drho values}). 
Similar considerations apply to adaptive mesh refinement simulations which typically have high resolution in the highest density regions in the galaxy disk, but not in the CGM.

\section{Summary and Future Work}\label{sec: conclusions}

In this study we develop a new method to analyze ionic column densities measured in the CGM, assuming that the cool ($T\sim10^4\K$) photoionized CGM spans a large dynamical range in gas density, and that the small high-density clouds are hierarchically embedded in larger low-density clouds.  
Our new method utilizes the formalism of the Absorption Measure Distribution (AMD), 
defined as the gas column per decade in gas density $\AMD$, which was originally developed for the analysis of `warm absorbers' near AGN. We demonstrate that this formalism allows combining (or `stacking') the information available from different objects
during the absorption line modeling, thus yielding significantly tighter constraints on CGM properties compared to traditional analysis methods which model each object individually.  

We apply our new method to the COS-Halos sample of low-redshift $\sim L_*$ galaxies, and find the following:
\begin{enumerate}
\item The 624 ionic column measurements and limits in all 44 COS-Halos sightlines, from the low-ions (e.g.\ \mgiip, \oip)
  to \ovip, can all be fit with a single normalization and slope of the AMD, namely $\AMD = 10^{18.5}(\rho/20\rhobar)^{0.05}\cm^{-2}$. The AMD spans the density range $20\rhobar < \rho < \rhomax$, 
 where $\rhobar$ is the cosmic mean baryon density 
 and $\rhomax$ is a maximum density fit separately to each object. 
 This success of our new fitting method supports our assumption that the CGM density structure is hierarchical. 
  We use cross-validation to demonstrate that the new fitting method is superior to traditional constant-density methods used in the literature, in terms of its ability to predict unseen data.
    
 \item Our fit provides $\rhomax$ of each sightline in the sample, which can be used
  to infer the covering factor of gas as a function of $\rho$. 
  The covering factor decreases roughly logarithmically with increasing $\rho$, from $90\%$ for
   $\rho\approx50\rhobar$ to $10\%$ for $\rho\approx5\times10^5\rhobar$.  

 \item Our results suggest that $\deriv\Mcool/\deriv(\log\rho) \propto \rho^{-0.2}$, i.e.\ a roughly equal cool CGM mass per decade in $\rho$. The total mass is $\Mcool=(1.3\pm0.4)\times10^{10}\msun$, a factor of five lower than estimates based on constant-density modeling. The derived $\Mcool$ is only $\sim$5\% of $(\Omega_{\rm b} / \Omega_{\rm DM})M_{\rm halo}$, the cosmic baryon budget of an $L_*$ galaxy.

 \item Since $\AMD \sim \nH \rc(\rho)$, where $\rc(\rho)$ is the characteristic size of clouds with density $\rho$, the flat slope of the AMD implies that $\rc(\rho)$ scales as $\sim \rho^{-1}$. This scaling implies that clouds in the CGM span a large dynamical range in size, from $\rc(50\rhobar)\approx35\kpc$ of the low density \ovip-phase to 
$\rc(5\times10^5\rhobar)\approx6\pc$ of the densest phase. This result can be tested by measuring the coherence scale of different ions with multiple sightlines towards lensed and binary quasars. 

 \item Based on the fit $\rhomax$ distribution as a function of impact parameter, we find an average cool baryon density profile of $\rhocool(R) =  97 (R/\rvir)^{-0.97} \msun \kpc^{-3}$ at $20\kpc<R<300\kpc$, where $R$ is the radial distance from the galaxy center. The uncertainty on the index of the profile is $\pm0.31$. This profile is significantly flatter than the dark matter profile at the same scales.

 \item Extrapolating the best-fit $\AMD$ down to $\rho=8\rhobar$ correctly predicts the \neviiip\ column observed in the CGM of $\sim L^*$ galaxies at $z\sim 0.7$, which may suggest that \neviiip\ absorption also originates in photoionized gas. 

\item The large range in densities found here coupled with the small range in temperature of photoionized gas  ($6000<T<30\,000\K$) together imply that the gas pressure increases strongly with density. Self-gravity and radiation pressure are too weak to establish this pressure gradient. The nature of the physical mechanism which generates the deduced density structure is currently an open question. 

\end{enumerate}
The results of this work can be expanded to find the dependence of the cool gas density structure on galaxy
star-formation rate, by applying the methodology independently to the 29 star forming galaxies and 15 quiescent galaxies in the COS-Halos sample. The low detection rate of $\novi$ in red galaxies compared to the high detection rate in blue galaxies (\citealt{Tumlinson+11}) might suggest that the two galaxy types differ in their cool gas density structure, although our results suggest these differences are not huge given the overall success of the universal fit. If the CGM of more massive quiescent galaxies is filled with a hot ($T\gtrsim10^6\K$) shock-heated gas phase, which is weak or absent in less massive star-forming galaxies (see e.g.\ \citealt{BirnboimDekel03,Keres+05,DekelBirnboim06}), 
then the high pressure from the hot plasma could compress the lowest density cool gas clouds, effectively suppressing the $\rho = 50\rhobar$ gas required to produce \ovip\ (see \S\ref{sec: hot phase}). 
We will explore this interesting possibility in future work.

Our analysis can also be expanded to study the density structure of low-mass halos and halos at high redshift, 
by applying the same methodology to the COS-Dwarfs sample of low-luminosity galaxies (\citealt{Bordoloi+14}), 
and to galaxy-selected absorption samples at $2<z<3$
(e.g.\ \citealt{Hennawi+06a,Crighton+11,Rudie+12, Lau+15}). 
Additional constraints can be deduced from analyzing absorption-selected samples such as 
KODIAQ (\citealt{OMeara+15}) and CASBaH (\citealt{Meiring+13}) in the context of the hierarchical picture, under the assumption that these absorbers are associated with galaxies. 

Further constraints on CGM properties and tests of our results can be derived from analyzing the gas kinematics in the context of the hierarchical picture (\S\ref{sec: kinematics}),
and from measurements of the coherence scale of different ions using sightlines with small transverse separations (\S\ref{sec: coherence}). These are all important topics for future work.

\acknowledgements

JS wishes to dedicate this work to the memory of his father, Izhak Z.~Stern, who passed away while this manuscript was being prepared. Izhak was an engineer by profession whose capacity for analytical and critical thinking inspired JS to pursue a career in science. 

The authors wish to thank David Hogg for proposing the cross-validation technique, and Avishai Dekel, Ehud Behar, Amiel Sternberg, Eve Ostriker, Ari Laor, Neil Crighton, Jose O{\~n}orbe, and Fred Davies for useful discussions. We also wish to thank the anonymous referee for careful reading of the manuscript and insightful comments. 
JS acknowledges financial support from the Alexander von Humboldt foundation.

\bibliographystyle{apj}

\appendix

\section{The relation between the discrete and continuous models}\label{app: discretization}

We calculate the relation between the the fine sampling of the gas densities calculated by \cloudy, which span  $\rhomin < \rho<\rhomax$ with $\rhomin=20\rhobar$, and the discrete densities $\rhoi$ used in \S\ref{sec: CGM characteristics} and Figures~\ref{fig: realization projected}--\ref{fig: realization slice}. 
Approximating the density gradient calculated by \cloudy\ as continuous implies
\begin{equation}\label{eq: rhoi}
 \rhoi = \frac{\int_{10^i \rhomin} ^{10^{i+1}\rhomin} \rho\deriv x}{\int_{10^i \rhomin} ^{10^{i+1}\rhomin} \deriv x} 
       = \frac{\mp}{X}\cdot\frac{\NH'(10^{i+1}\rhomin)-\NH'(10^{i}\rhomin)}{x(10^{i+1}\rhomin)-x(10^{i}\rhomin)} ~,
\end{equation}
where $\mp/X\approx1.4\mp$ is the gas mass per hydrogen atom and $x$ is the depth measured from the outer layer. 
The value of $\NH'$ can be derived from eqn.~(\ref{eq: NH'}).
The value of $x$ can be derived from the AMD by noting that $\AMD = \nH\deriv x /\deriv\log \rho$, which together with the power-law parameterization of the AMD (eqn.~\ref{eq: AMD}) yields
\begin{equation}
 x(\rho) = \frac{\mp}{X\rhoz}\cdot\frac{\dNz}{\ln 10} \cdot \frac{1}{\beta-1} \cdot \left[\left(\frac{\rho}{\rhoz}\right)^{\beta-1}-1\right]
\end{equation}
Using the best-fit $\dNz=10^{18.54}\cm^{-2}$ and $\beta=0.05$ from eqn.~(\ref{eq: AMD fit}), eqn.~(\ref{eq: rhoi}) implies that $\rhoi =2.4\times10^i\rhomin=48\times10^i \rhobar$.

\end{document}